\documentclass[conference]{IEEEtran}
\IEEEoverridecommandlockouts
\usepackage{cite}
\usepackage{amsmath,amssymb,amsfonts}
\usepackage{algorithmic}
\usepackage{graphicx}
\usepackage{textcomp}
\usepackage{xcolor}
\def\BibTeX{{\rm B\kern-.05em{\sc i\kern-.025em b}\kern-.08em
    T\kern-.1667em\lower.7ex\hbox{E}\kern-.125emX}}

\usepackage{booktabs}
\usepackage{url}
\usepackage{array}
\usepackage{longtable}
\usepackage[hidelinks]{hyperref}

\usepackage{enumitem}      
\usepackage{float}         
\usepackage{ragged2e}      
\usepackage{tikz}          
\usetikzlibrary{positioning,arrows.meta,fit}
\newcommand{\dsname}{SpecMine}
\newcommand{\num}[1]{\mbox{#1}}

\newcommand{\src}[1]{{\normalfont\footnotesize~\mbox{[#1]}}}

\begin{document}

\title{Spec-Driven Development as Prescribed and as Practiced}

\author{\IEEEauthorblockN{Shyam Agarwal\IEEEauthorrefmark{1}\IEEEauthorrefmark{3},
Anmol Singhal\IEEEauthorrefmark{1}\IEEEauthorrefmark{3},
Andrew Macvean\IEEEauthorrefmark{2},
Bogdan Vasilescu\IEEEauthorrefmark{1},
Travis Breaux\IEEEauthorrefmark{1}}
\IEEEauthorblockA{\IEEEauthorrefmark{1}Carnegie Mellon University}
\IEEEauthorblockA{\IEEEauthorrefmark{2}Google}
\IEEEauthorblockA{anmolsinghal@cmu.edu, shyagarwal@cmu.edu, amacvean@google.com, vasilescu@cmu.edu, breaux@cs.cmu.edu}
\IEEEauthorblockA{\IEEEauthorrefmark{3}Both authors contributed equally to this research.}}

\maketitle

\begin{abstract}
  AI coding agents implement software from natural-language instructions, but the intent behind that work is lost when stated only in chat sessions. Spec-driven development (SDD) addresses this by having developers and agents write specifications that guide implementation, yet evidence on its use comes largely from tool vendors. We study SDD in practice through a thematic analysis of Reddit discussions, interviews with 21 practitioners, including eight tool makers, and 266 active and popular repositories in SpecMine, a corpus of 470,795 spec files from 73,030 public GitHub repositories that we release. Practitioners agreed that specs should capture intent, but many also included implementation details, used SDD selectively and mainly for new features, and reported more reliable agent behavior alongside greater planning and review costs. Our findings show that implementation detail may help agents while making specs prone to decay, that tool-specific record structures shape what people and agents see, and that participants described spec--code checks mainly when changes are completed rather than over time. We also find that individual benefits may come from clarifying intent before delegation, while team benefits may depend on SDD becoming a shared process. We derive implications for tool builders, practitioners, and researchers. 
\end{abstract}


\section{Introduction}\label{sec:intro}

AI coding agents such as Cursor, Claude Code, and
OpenHands\footnote{\href{https://cursor.com}{Cursor}, \href{https://claude.com/product/claude-code}{Claude Code}, \href{https://openhands.dev}{OpenHands}} can plan, modify, test, and debug code across a
repository with little human intervention~\cite{yao2023react}. Developers
increasingly delegate implementation to an agent and intervene to refine
instructions, review results, or correct failures~\cite{Cynthia_2026}.
Delegation raises the question of how to communicate intent to an agent. In
\emph{vibe coding}~\cite{meske2025}, a developer states intent in a chat
session and iterates on the agent's output, and the intent is lost with the
session: teammates cannot review it, and future agents cannot read it. Gains
in individual productivity may then not carry over to teams, where
coordination, review, and maintenance costs offset
them~\cite{Abraho2025SoftwareEB}, and review capacity and system stability
degrade as agents produce more code~\cite{diaz2026sddase}.

\begin{figure}[t]
  \centering
  \includegraphics[width=0.65\columnwidth]{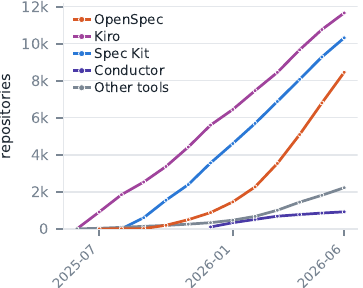}
  \caption{Cumulative GitHub repositories per named tool, by first commit to a spec
    file. Recent agentic SDD tools are spreading fast. Excludes
  \texttt{caffeine.ai}, a generator account emitting one repository per app (13,749).}
  \label{fig:adoption}
\end{figure}
Spec-driven development (SDD) answers this problem in agentic software
engineering by making an explicit specification of intended behavior the
artifact from which implementation proceeds. Specifications (specs) are
written by developers or agents, typically in Markdown, and agents read them
to plan, implement, and verify. SDD has spread quickly. GitHub, Amazon, and
Fission AI produced dedicated tools (Spec~Kit~\cite{speckit},
Kiro~\cite{kiro}, OpenSpec~\cite{openspec}), practitioner communities have
formed around them, and we found spec files in over 73,000 public GitHub repositories, almost all first committed in 2025 or later
(Figure~\ref{fig:adoption}). SDD is also an old idea. Software engineering has
held for five decades that a specification should precede and govern
implementation, from the waterfall model~\cite{royce1970waterfall} through
the spiral model~\cite{boehm1988spiral} and Scrum~\cite{schwaber2001agile} to
the requirements engineering standard ISO/IEC/IEEE 29148~\cite{iso29148}, and
requirements engineering has studied for as long what makes the
practice hard to sustain: specs are costly to write, go stale as systems
evolve, and are hard to trace to the code that satisfies
them~\cite{gotel1994traceability}. The idea has returned because the reader
has changed. An agent consumes natural language directly, so a spec no longer
has to be formal, executable, or bound to a generator. It can be drafted in
minutes, in conversation with the agent that will implement it, and the cost
that kept specification-first development out of most projects may no longer
apply.

Despite this traction, no empirical study has broadly examined the modern
practice. What is known comes from the people who advocate SDD.
Tool documentation, gray literature, and early research prescribe workflows
and promise better quality, less agent drift, and higher productivity.
However, their evaluations rarely extend beyond the authors' own projects
(Section~\ref{sec:background}). The pre-AI history gives reasons to expect
specific trouble. Cost, decay, and weak traceability have limited earlier
specification-first
practices, and whether natural-language specs read by agents escape them is
unknown. Seyff and Glinz observe that the practice ``is largely silent on how the
specification is produced''~\cite{seyff2026sketches}; it is just as silent
on who produces it, what they put in it, who reads it, whether anyone keeps
it current, and whether promised benefits appear outside its makers'
demonstrations.

This paper begins to answer these questions with the first systematic study
of SDD as it is practiced. We organize a scattered body of vendor
documentation, community discourse, practitioner experience, and repository
evidence into a structured account of the paradigm, and we ask how that
account differs from what its proponents prescribe.

\begin{description}
  \item[RQ] How do practitioners practice SDD, and how does their practice
    differ from what SDD tools and their makers prescribe?
\end{description}

\noindent We divide this question into four sub-questions that follow the
lifecycle of a spec.

\begin{description}
  \item[RQ1] What content do practitioners put in spec files?
  \item[RQ2] What SDD workflows do practitioners follow, and what factors
    shape their choices?
  \item[RQ3] What benefits have practitioners observed as a consequence of adopting SDD?
  \item[RQ4] What practical challenges emerge from adopting SDD?
\end{description}

We answer these questions with a concurrent triangulation design
(Section~\ref{sec:method}), combining a content analysis of practitioner
discussions on Reddit, semi-structured interviews with 21 practitioners, and an
analysis of 266 active and popular repositories in SpecMine, a corpus of
470,795 spec files from 73,030 public GitHub repositories that we release.
The repository analysis tests the qualitative findings where repositories
make them observable, mainly on spec content and workflows; its sample is
dominated by OpenSpec and Spec~Kit (89\% of its files). We interviewed both \textit{tool makers},
who built and released SDD tools, and \textit{tool users}. The makers' accounts
largely match what the gray literature prescribes, so where the two groups
disagree, practice departs from prescription. Section~\ref{sec:discussion}
discusses what the findings imply for spec
detail, intent preservation, spec--code consistency, and SDD's benefits
to individuals and teams. This paper makes three contributions:
\begin{itemize}
  \item The first systematic characterization of SDD as practiced, covering
    spec content, workflows, benefits, and challenges, combining practitioner
    discourse and interviews with repository evidence, and distinguishing what
    tool makers prescribe from what tool users do;
  \item SpecMine, a public corpus of 470,795 spec files from 73,030 public GitHub repositories, with files attributed to 18 SDD tools by directory, plus a separate census of 98,574 Kiro artifacts, enriched with recorded commit histories and parsed
    document structure; and
  \item Implications for tool builders, practitioners, and researchers.
\end{itemize}

\section{Background and Related Work}\label{sec:background}

\textit{Specification before implementation.} Software engineering has
long held that a specification should precede and govern
implementation~\cite{royce1970waterfall,boehm1988spiral,schwaber2001agile,iso29148}.
Requirements engineering has studied what specifications are for and where
they fail~\cite{zave1997darkcorners,nuseibeh2001twinpeaks}, and how to make
them precise and checkable~\cite{mavin2009ears,lucassen2016qus,femmer2017smells}.
Specifications are expensive to write, go stale as systems evolve, and are
hard to relate to the code that satisfies them~\cite{gotel1994traceability}.
Approaches that bind spec and code include Design by
Contract~\cite{meyer1992}, an earlier \emph{Specification-Driven Development}
that treated tests and contracts as executable specs~\cite{ostroff2004},
model-driven engineering~\cite{diaz2026sddase,seyff2026sketches}, and
behavior-driven development~\cite{diaz2026sddase}. Contemporary SDD revives
the practice under new conditions. An AI coding agent consumes natural
language directly, so a spec need no longer be formal, executable, or bound
to a generator, and it can be drafted in minutes.
Whether its benefits materialize, and whether the
difficulties that limited earlier practice recur, has not been examined in
projects using SDD tools.

\textit{What SDD proponents prescribe.}\label{sec:bg-prescribe} Tools such
as GitHub Spec~Kit~\cite{speckit} and Amazon Kiro~\cite{kiro} prescribe a
sequence of Markdown documents, typically a spec, a plan or design, and a
task list, written before implementation. Their documentation claims that SDD
improves quality, makes outcomes predictable, or speeds up delivery~\cite{speckit,kiro,openspec}. Appendix~\ref{app:tools} compares what five of these tools prescribe.
Practitioner accounts classify SDD by how long a spec is kept and how
strictly it governs the code~\cite{bockeler2025,piskala2026sdd}. Other work
frames SDD as a discipline for agentic software engineering in
teams~\cite{diaz2026sddase}, extends it to sketch-based
modeling~\cite{seyff2026sketches}, security~\cite{marri2026constitutional},
control of spec--code drift~\cite{grabowski2026specgrowth}, and autonomous
operation~\cite{roy2026kitchenloop}, or offers guides for adopting it in
small teams~\cite{crider2026aidelivery}. None of these works studies SDD use
beyond the authors' own projects.

\textit{Empirical evidence on SDD.}\label{sec:bg-evidence} Evidence for
the claims of SDD proponents is limited~\cite{diaz2026sddase}. Hill's study
of 88,052 pull requests~\cite{hill2026} operationalizes SDD at the
pull-request level, through indicators such as linked issues, references to
specification or design documents, and structured requirements in PR
descriptions, rather than tool-generated spec files such as
\texttt{spec.md}, so its results are not directly comparable to ours.
Other studies are small: a four-project workflow study~\cite{santos2026ddd},
a course study with 14 students~\cite{tanaka2026}, and two frameworks
their authors evaluated on their own
systems~\cite{marri2026constitutional,roy2026kitchenloop}.

\section{Research Method}\label{sec:method}

We follow a concurrent triangulation strategy~\cite{easterbrook2008selecting}:
we collected and analyzed three sources largely in parallel and use each to
confirm, qualify, or contradict the others. Two sources record what
practitioners say: a thematic analysis of 154 Reddit posts and comments (99
relevant to SDD practice) and semi-structured interviews with 21
practitioners, which together address all four RQs. The third records what
practitioners do: a repository sample of 266 active and popular repositories
drawn from SpecMine (Section~\ref{sec:method-mining}), whose files, commit
histories, and pull requests show what specs contain, how they are scoped and
retained, how often they are revised, and who writes and comments on them.
Wherever a qualitative finding makes a claim that is observable in
repositories, we test it against SpecMine: what specs contain and how they
are scoped and retained (RQ1), how they are revised and reviewed (RQ2), and
who writes them (RQ3, Finding~6).
The repository sample does not measure effort, perceptions, or reasons, so the remaining
findings on benefits and all findings on challenges (RQ3 and RQ4) rest on the
qualitative sources alone. The three
analyses were conducted concurrently, with one dependency: a formative pass
over a first batch of Reddit records preceded the interviews and shaped their
protocol (Section~\ref{sec:method-reddit}). The proper Reddit analysis, which
recoded those records alongside a second batch, then proceeded in parallel
with the interviews and the repository analysis, and its codes were merged with
the interview codes into one codebook (Section~\ref{sec:method-merge}). Where
the sources disagree, we report both and interpret the disagreement
(Findings~3 and~4, Section~\ref{sec:disc-record}). We also read the documentation of the five dedicated SDD tools in the repository
sample, as it stood when collection ended (Appendix~\ref{app:tools}). We did not code
this documentation. We use it only to check the qualitative and repository
findings against what the tools prescribe.

\textit{Trustworthiness.} We followed Lincoln and Guba's criteria for
qualitative research~\cite{lincoln1985naturalistic}. For credibility, we
triangulated across the three sources, had the first co-lead code the interviews and
the first batch of Reddit records and the second co-lead independently code the second batch,
refined the codebook in their discussion and then in three meetings
with all co-authors, and re-coded all transcripts and Reddit records against
the merged codebook before formulating claims. We tracked saturation as
interviews accrued. For dependability and confirmability, the interview
protocol, the keyword list, and the codebook are in the replication
package, and the findings quote participants and Reddit records so
claims can be traced to their sources. As is common in interpretive thematic
analysis, we did not compute inter-rater
agreement~\cite{mcdonald2019reliability} (consequences in
Section~\ref{sec:threats}).

\textit{Ethical considerations.} Carnegie Mellon University's Institutional Review
Board approved the study. Interview participants gave verbal consent to
recording, could decline recording, skip questions, or stop at any time, and
were not compensated. Recordings were deleted after transcription, and
transcripts were de-identified by removing employer, product, and project
names; because several participants build publicly known SDD tools, we do
not release transcripts and report demographics only in aggregate, under
pseudonyms (P1--P21). SpecMine is drawn from public repositories: each
record retains its repository's license, the release keeps public author
logins but drops commit emails and profile fields, and we report results in
aggregate without identifying contributors.

\subsection{Reddit Analysis}\label{sec:method-reddit}

Reddit hosts active communities around AI coding tools and SDD frameworks in
which practitioners describe their workflows, report problems, and debate the
value of specs, and prior work has used it as a source of practitioner
discourse in SE~\cite{iqbal2021reddit,popoola2025interns}.
We analyze developers' self-reported experiences in these discussions to
answer the RQs. A formative pass over a first batch of records informed the
interview protocol (Section~\ref{sec:method-interviews}); the full analysis
ran concurrently with the interviews.

\textit{Data collection.} We collected posts and comments from Reddit manually 
by using its web search interface, across 23 communities in
four groups: SDD frameworks (e.g., r/BMAD\_Method,
r/SpecDrivenDevelopment, r/speckit, r/openspec, r/kiroIDE); AI coding agents
(e.g., r/ClaudeCode, r/cursor, r/GithubCopilot, r/codex); AI-assisted
development in general (e.g., r/vibecoding, r/ChatGPTCoding, r/AI\_Agents);
and general SE (e.g., r/ExperiencedDevs, r/programming,
r/softwaredevelopment). Within each community, we used 70 search terms
drawn from gray literature on SDD and agentic planning (e.g.,
``spec-driven,'' ``spec.md,'' ``PRD,'' ``task decomposition,''
``CLAUDE.md'') to find relevant posts and comments. We restricted each query to that community, 
collected posts and comments dated February 2025 to September 2026, and sorted by Reddit's
\emph{Top} ranking, so that records are ordered by engagement within their own
community. This process yielded 1{,}610 searches. We recorded every result meeting the
criteria below together with the query that surfaced it, and stopped a search
once two consecutive result pages produced nothing new that qualified.
Appendix~\ref{app:reddit} gives the records found per community.

\textit{Filtering and sampling.}\label{sec:filterproc} We retained records
with at least two upvotes and 100 characters, excluding deleted or removed
content and automated authors, and judged on topic for SDD practice. Records
were deduplicated by permalink, since one record is surfaced by many queries;
we kept the number of distinct queries that surfaced each record as a measure
of how densely it matches the vocabulary of SDD. Two authors searched
independently and reconciled their results in three rounds, keeping at each
round the 30 highest-ranked posts and 30 highest-ranked comments per
community, so that large communities could not dominate, yielding
\num{1{,}497} records (Appendix~\ref{app:reddit}).

\textit{Thematic analysis.} We analyzed the filtered records in two stages
using inductive, iterative thematic analysis~\cite{braun2006thematic}. In the
\emph{formative} stage, the first co-lead open-coded the 100 highest-ranked
records from the first data collection (62 posts and 38 comments dated
February 2025 to April 2026, 79 of them relevant to SDD practice) into seven
high-level codes (authorship, workflow, evaluation, maintenance, tooling,
benefits, and challenges) and 138 low-level codes that refined them; benefits,
for example, included shipping velocity, reduced cognitive load, engineering
quality, and maintainability. These codes shaped the interview protocol
(Section~\ref{sec:method-interviews}) but did not reach saturation by
themselves. In the \emph{second-coder} stage, conducted while the interviews were under
way, the second co-lead independently open-coded the 101 highest-ranked posts,
62 of them relevant to SDD practice,
into codes covering practices, tools, outcomes, and challenges. Because 47
records appeared in both samples, the merged analysis uses the second co-lead's
coding of those and the formative coding of the other 53 (37 relevant).

The two coders then merged their codes through a
series of discussions: each code was compared with the other set's codes and
the records it had been applied to, codes describing the same concept
were merged and the others kept, and disagreements were resolved in
discussion. The formative stage contributed 73 codes from its 53 retained
records and the second-coder stage 125, of which 42 were shared, giving a
Reddit codebook of 156 low-level codes under 14 high-level codes over 154
records, 99 of them relevant to SDD practice. We did not code further
because, once merged with the interview codes
(Section~\ref{sec:method-merge}), all but 15 Reddit codes also emerged in
the interviews, fewer than 4\% of the final codebook. The keyword list and
the merged codebook are in the replication package.

\subsection{Semi-Structured Interviews}\label{sec:method-interviews}

\textit{Protocol.} We derived the interview questions from the formative
Reddit analysis: its seven high-level codes shaped the main questions, and
its low-level codes the follow-up probes. The interviewer probed a theme only
once a participant raised it, so the themes reported in
Section~\ref{sec:findings} emerged from the interviews rather than from this
preliminary set. The protocol has 17 main questions in four sections:
background and context, SDD workflows in practice (RQ1--RQ2), benefits (RQ3),
and challenges (RQ4). For participants unfamiliar with the term, the
interviewer defined SDD as ``the practice of creating structured,
behavior-oriented documents that you use as context when working with AI
coding agents,'' adapted from B{\"o}ckeler~\cite{bockeler2025}. All authors
reviewed the protocol before the first interview; it is available in the replication package.

\textit{Recruitment.} We recruited participants by (1) posting a
recruitment call on LinkedIn and in the subreddits of the Reddit corpus; (2)
e-mailing and direct-messaging Reddit users who made substantive
contributions to SDD discussions; and (3) contacting contributors to
open-source projects on GitHub with at least 100 stars whose repositories
contained spec files (Section~\ref{sec:method-mining}). We contacted 29
practitioners on LinkedIn, 59 on Reddit, and 35 on GitHub, and received 12
responses to our public announcements. Interested practitioners completed a
screening survey: participants had to be at least 18 years old, fluent in
English, have at least one year of software development experience, have
used AI coding agents, and have used SDD. This process yielded 21 interviews.

We assign each participant an \textit{SDD role}: \textit{tool makers}
created an SDD tool and released it publicly, and \emph{tool users} are all
others, including those who use tools built by others and those who follow
their own workflow with general-purpose agents. We interviewed tool makers
even though their tools are documented in gray literature, because
documentation describes what a tool prescribes but not how its makers apply
it or why; we treat their accounts as the prescribed view against which we
compare tool users' practice (Section~\ref{sec:intro}).
Table~\ref{tab:participants} summarizes demographics.

\begin{table*}[t]
  \caption{Participant demographics (N = 21).}
  \label{tab:participants}
  \footnotesize
  \begin{tabular}{@{}l p{0.78\textwidth}@{}}
    \toprule
    Dimension & Participants \\
    \midrule
    SDD role & Tool maker (8), tool user (13) \\
    Years of experience & $<$10 (6), 10--19 (7), 20--29 (5), $\geq$30 (3) \\
    Primary role & Software developer (11), founder, manager, or product
    manager (6), architect or consultant (3), researcher (1) \\
    Current SDD tool & OpenSpec or derivative (7), Spec Kit or derivative (3),
    Superpowers (3), own tool (3), no dedicated tool (3), Kiro (1), BMAD (1) \\
    \bottomrule
  \end{tabular}
\end{table*}

\textit{Data collection and analysis.} The first co-lead conducted all
interviews via Zoom between May and September 2026; they lasted 39--106
minutes (median 55). Zoom transcribed the recordings, the transcripts were
de-identified, \texttt{gpt-5.4-mini} corrected grammar and speech
disfluencies, and the first co-lead checked the corrected transcripts against
the originals. The first co-lead then analyzed the transcripts using
inductive, iterative thematic analysis~\cite{braun2006thematic}, open coding
segments against the RQs and naming each code \emph{high-level code:
low-level code}, with the seven formative codes as the starting high-level
codes. The interview analysis ran in parallel to the second-coder
analysis (Section~\ref{sec:method-reddit}), each by one
author in their own wording, and yielded 412 low-level codes.

\textit{Merging Reddit and interview codes.}\label{sec:method-merge} Two
authors merged the interview codes with the Reddit codebook
(Section~\ref{sec:method-reddit}) by the same procedure: each Reddit code was
compared with the interview codes and merged into one when both described
the same concept or kept as a new code otherwise, with every merge reviewed
and disagreements resolved via discussion. The Reddit
codebook held 156 low-level codes and the interviews 412; after merging, the
codebook held 427 low-level codes under 15 high-level codes
(Appendix~\ref{app:dimensions}). In three one-hour meetings, all authors
then grouped the merged codes into themes, identified relationships between
them, and formulated the claims reported in Section~\ref{sec:findings}. Once
the codebook stabilized, the first co-lead and the second co-lead re-coded all transcripts and Reddit records against it for consistency, resolving uncertain codes with the other authors. After this re-coding, the 99 Reddit records relevant to SDD practice carry 174 of the 427 codes.

To assess saturation, we tracked the new codes each interview added to the
Reddit codebook, in interview order, and considered
saturation reached once three consecutive interviews added at most
5\% new codes~\cite{guest2020saturation}. This criterion was met from the
seventh interview onward, and 95\% of the codes had emerged by the 18th
interview. Codes on workflows (RQ2) and challenges (RQ4) accrued later than
those on spec content (RQ1) and benefits (RQ3). The codebook is in the replication package.

\subsection{Repository Mining}\label{sec:method-mining}

Repositories record the artifacts that the practice described on Reddit and
in interviews produced. We built \emph{SpecMine}, a corpus of spec files
published on GitHub with their commit histories, the tool that produced each
file, and the pull requests that modified it, and use it to analyze spec
content, workflow traces, and authorship for
RQ1, RQ2, and Finding~6
(details in Appendix~\ref{app:mining}).

\textit{Data collection and tool attribution.} SDD tools are not
identifiable from repository metadata, so we identified repositories by the
files they contain. Most tools write a file named \texttt{spec.md} or
\texttt{specs.md}, which we used as the search filter. Kiro instead writes
\texttt{requirements.md}, \texttt{design.md}, and \texttt{tasks.md} under
\texttt{.kiro/specs/}, so we searched for that directory separately and
treat Kiro as a distinct census throughout. We queried GitHub's code-search
and REST APIs in July 2026, and excluded files held in package caches or
vendored dependency trees and files first committed before 2024, which
predate SDD practice. The corpus holds 470,795 specs across 73,030
repositories and 44,521 owners, and 98,574 Kiro artifacts across 12,910
repositories; 99.7\% of specs were first committed in 2025 or later and
92\% in 2026. Because Kiro's \texttt{requirements.md} is the artifact that
corresponds to a \texttt{spec.md} elsewhere, we treat its 32,972 requirements documents as specs. Together with the 470,795 spec files, they give 503,767 specs in 85,300 repositories, because 227 repositories appear in both searches.

We attribute each spec to a tool from the directory that tool creates, such
as \texttt{openspec/} for OpenSpec or \texttt{.specify/} for Spec~Kit. The
patterns identify 18 named tools; a file that matches none records that no
tool marker was found, not that no tool was used. Where a tool's layout
separates artifacts meant to persist from those treated as transient, the
same patterns assign a \emph{lifecycle role}. Only OpenSpec does so. It keeps one spec for each \emph{capability}, its term for a feature or other part of the system, under \texttt{openspec/specs/}. We call these \emph{capability specs}. It writes each change as change specs under \texttt{openspec/changes/} and moves them to \texttt{changes/archive/} when the change is complete. We therefore report roles for OpenSpec only.

\textit{Repository sample.} A corpus-wide analysis would mix maintained
projects with homework, template clones, and one-off experiments.
We therefore analyze one filtered sample throughout, built with five conjunctive filters
(Table~\ref{tab:corpus}): attribution to a named tool, so that every analyzed
spec has known provenance; at least ten stars and a recognizable open source
(SPDX) license, to reduce the influence of toy and unredistributable
repositories; public, not archived, and not a fork; and at least ten distinct
commits touching the repository's specs, to separate sustained SDD
practice from adoption that did not continue (29.9\% of named-tool
repositories above the star threshold have at most one spec commit). The
repository sample holds 266 repositories and 36,890 files,
including 13 Kiro repositories with 654 requirements
documents.\label{sec:method-mining-analysis} For each spec we record its commit
history, its lifecycle role where the tool encodes one, and 39 structural
features parsed from its text. We also collect the pull requests (PRs) GitHub associates with each spec commit. Review is observable only for specs that pass through a PR, because some repositories commit specs directly to a branch. A broader crawl of 949 repositories (Appendix~\ref{app:mining}) yields 5,979 PRs from 581 repositories, and 81.2\% of these PRs also modify code. Tool adoption,
dated by the first commit touching a spec attributable to a tool, spans
August 2025 to July 2026 and is uneven: twelve of the 18 tools survive the
filters, and OpenSpec (165 adoptions) and Spec~Kit (71) account for 85\% of the
279 repository--tool adoptions and 89\% of the sample's files
(Figure~\ref{fig:adoption}; Figure~\ref{fig:timeline} in
Appendix~\ref{app:mining} adds specs per month). Findings reported
across tools thus rest disproportionately on OpenSpec.

\begin{table*}[t]
  \caption{The SpecMine corpus (left) and the repository sample by successive
  filter (right; spec files and Kiro requirements documents counted together). Repository counts include 13,749 repositories generated by the \texttt{caffeine.ai} account.}
  \label{tab:corpus}
  \footnotesize
  \begin{tabular}[t]{@{}lr@{}}
    \toprule
    Corpus & Count \\
    \midrule
    Specification files & 470,795 \\
    \quad repositories / owners & 73,030 / 44,521 \\
    \quad named SDD tools & 18 \\
    Kiro artifacts, all kinds & 98,574 \\
    \quad repositories & 12,910 \\
    Specification-file commits & 780,335 \\
    Repositories with a recognized license & 28,698 \\
    \bottomrule
  \end{tabular}
  \hfill
  \begin{tabular}[t]{@{}lr@{}}
    \toprule
    Repository sample & Repos / files \\
    \midrule
    All kept specifications & 85,300 / 503,767 \\
    Attributed to a named tool & 48,732 / 396,818 \\
    $\geq$10 stars & 1,446 / 50,858 \\
    Public, not archived, not a fork & 1,444 / 50,836 \\
    Recognized SPDX license & 1,031 / 42,813 \\
    \textbf{$\geq$10 distinct specification commits} & \textbf{266 / 36,890} \\
    Specification commits in the sample & 13,640 \\
    \bottomrule
  \end{tabular}
\end{table*}

\textit{Structure (RQ1).} We parse each file's Markdown at its last
commit, normalize the headings (Appendix~\ref{app:headings}), and count each
once per file.
Files are grouped by tool and, for OpenSpec, by lifecycle role. We label a
section as core to a type when half its files contain it.

\textit{Content (RQ1, RQ2).} We used an LLM (\texttt{gemini-2.5-pro}, temperature
zero) to judge a stratified sample of 1{,}000 documents, drawn equally across
the 13 types and weighted by each type's number of files, on whether a spec states
\emph{requirements} (what the software must do and when the result is
correct), whether it also states \emph{implementation} detail (how it is to
be built), and what kind of work it describes, using the purpose and object
labels of Watanabe et al.~\cite{watanabe2025agentic}, with multiple labels
allowed. The two definitions come from requirements engineering rather than
from our interviews, so the measure is independent of the account it tests.
The model sees only the document text and must quote verbatim support for
every label (prompts in Appendix~\ref{app:prompts}). The 13 types cover every file in the sample except the 83 from Trae,
Claude Code, and Cursor, which had too few files to sample. The estimates
include the 31 documents that the judge marked as not specs. LLM-as-a-Judge validation details are in Appendix~\ref{app:mining}.

\textit{Revision (RQ2).} We count the distinct commits touching each file,
by document type.

\textit{Authorship and review (RQ2, RQ3).} We use commit authorship (to spec
files) and inline PR comments as proxies for measuring participation. Authors
are counted per file and per repository and labeled as the repository owner,
one of its three most frequent committers, another contributor,
or absent; comments are the inline PR comments on each spec file.

\section{Study Findings}\label{sec:findings}

We organize the findings by research question across the Reddit and interview analyses, and the repository sample
(266 repositories), which adds complementary evidence
on spec content and workflows (RQ1 and RQ2) and on spec authorship
(Finding~6). Quotes are from Reddit records R1--R99 and from interview participants P1--P21. Each finding header lists the sources that support the finding as R for Reddit, I for interviews, and S for the repository sample. Comparisons between tool
makers and tool users draw only on the interviews, because the Reddit data does not reveal whether authors build SDD tools.
Appendix~\ref{app:codes} lists the participants who support each claim in the findings,
Appendix~\ref{app:generic} summarizes shared practices, and
Section~\ref{sec:discussion} considers the implications.

\subsection{RQ1: What Practitioners Include in Spec Files}\label{sec:findings-rq1}

\textit{Finding 1: Practitioners agree that a spec must capture what
  to build; most tool users also include how to build it, especially when the
  task is complex or requires knowledge the agent lacks. In public repositories,
most specs include implementation detail, under sections their tool supplies.}\src{I, S}

Participants distinguished \emph{what} to build from \emph{how} to build it
(Table~\ref{tab:what-how}). The \emph{what} states the purpose of a spec change
and the system behavior the change should produce. We include verification-related information, such as acceptance criteria, as part of \emph{what} because most 
participants described this information as the conditions for accepting the resulting system behavior after the spec is implemented. The \emph{how} ranges from architecture and design to tasks and
code-level details such as data structures and library versions.

\begin{table*}[t]
  \caption{Taxonomy of \emph{what} and \emph{how} content in spec files.}
  \label{tab:what-how}
  \footnotesize
  \begin{tabular*}{\textwidth}{@{}l l @{\extracolsep{\fill}} l@{}}
    \toprule
    & Content type & Description (examples from participants) \\
    \midrule
    \emph{What} & Goals and intent & Problem statement, purpose, and
    rationale of a change \\
    & Requirements & User stories, user journeys, and use cases with flows
    and conditions \\
    & Acceptance & Acceptance criteria, edge cases, failure scenarios, and
    how to verify the result \\
    & Constraints & Non-functional requirements, guardrails (``no
    shortcuts''), out-of-scope items \\
    & Context & Business context and stakeholder needs \\
    \midrule
    \emph{How} & Design level & Architecture, components, interfaces, APIs,
    design decisions and trade-offs \\
    & Plan level & Task breakdown and sequencing, technology choices,
    database tables, APIs \\
    & Code level & Implementation approach, classes, data structures,
    algorithms, library versions \\
    & Cross-cutting & Coding conventions, domain knowledge, reference docs
    for unfamiliar libraries \\
    \bottomrule
  \end{tabular*}
  \vspace{-2mm}
\end{table*}

\textbf{Tool users put the \emph{how} into specs.} 
Most tool users routinely included
implementation-level content in the spec irrespective of the task; other users included such content depending on how much they trust the agent to accurately perform complex tasks. They added detail as tasks grew more complex or exceeded
what they believed the agent could do: P10's team specifies ``the how if we
believe it's in a place where AI has very low capability of doing it.'' They
also added knowledge the agent could not infer, such as internal APIs
and integrations, domain rules, and team conventions: P16's design documents
include API error codes and OpenAPI definitions ``so that it understands
better how the integration goes.'' 

\textbf{Tool makers keep the \emph{how} out of specs.} Most tool makers spoke about focusing specs on the ``what'' by ``defer[ring]
implementation for as long as you can'' (P2) and by excluding or separating the ``how''
into design and task documents, which P9 described as the ``top-level separation'' that SDD tools share. A few tool makers also included implementation detail in their specs (P13, P21). Most tool makers kept the structure of specs
consistent through templates, but some tool makers held that structure should vary, for example by keeping only
the problem definition fixed (P1) or by letting users remove template sections (P20).

\begin{figure*}[t]
  \centering
  \includegraphics[width=0.8\textwidth]{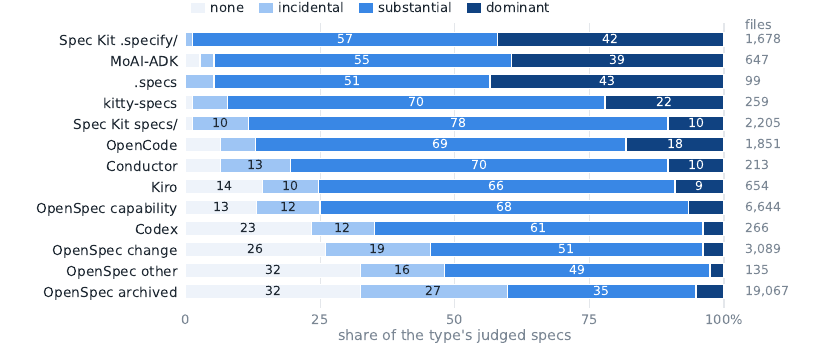}
  \caption{Share of specs describing implementation
    detail categorized by tool; corpus size at right. Scale is how much of
    a document that detail occupies: \emph{incidental} a line or two in
    passing, \emph{substantial} a section or several statements but not the bulk, \emph{dominant} most of the document.}
  \label{fig:content}
\end{figure*}

\begin{figure*}[t]
  \centering
  \includegraphics[width=0.8\textwidth]{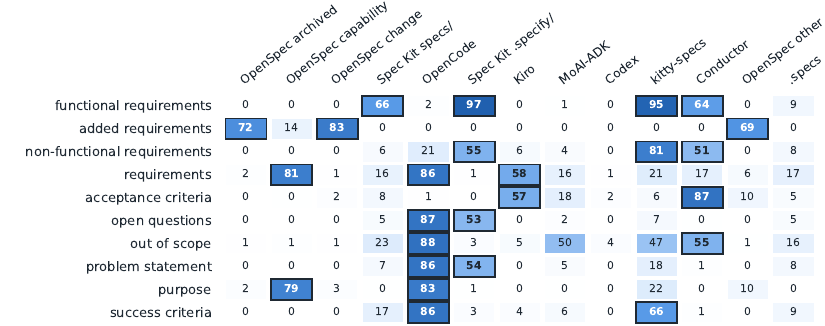}
  \caption{Sections heading a majority of files in more than one document type.}
  \label{fig:headings}
\end{figure*}

While nearly all specs in the repository sample that we analyzed using LLM-as-a-Judge include requirements, most spec files (77.6\%) contain implementation
detail, including 58.2\% wherein that material
occupies a section or more (Figure~\ref{fig:content}). We found that 75\% of such spec files
name classes, functions, file paths, data structures or library versions, and
71\% describe components, interfaces or data models. Tool makers describe
keeping implementation detail
out, and the tools they build expect separate files for spec, design and tasks (Appendix~\ref{app:tools}). OpenSpec and Spec~Kit also tell users explicitly to keep implementation detail out of the spec. This expectation contrasts with what we observed in most repositories.
The sections in committed specs largely match each tool's template: 81\% of OpenSpec capability specs head \emph{requirements}, 83\% of its change specs head \emph{added requirements}, and 87\% of Conductor files head \emph{acceptance criteria} (Figure~\ref{fig:headings}; Appendix~\ref{app:specmine-results}).

\subsection{RQ2: How Practitioners Work with Specs}\label{sec:findings-rq2}

Practices common to nearly all participants are summarized in
Appendix~\ref{app:generic}: agents drafted specs while
participants supplied the intent and reviewed the result, specs were refined
iteratively, and code was verified mainly through tests that agents wrote.
Appendix~\ref{app:generic} also describes the three types of workflow that participants
followed, which differ by where a human checks the agent's work. Below we report
where participants differed or practice departed from what tools prescribe.

\textit{Finding 2: Tool users mostly decide to use SDD by task size
and write specs mainly for new features.}\src{R, I, S}

Most practitioners used SDD selectively, and only a few used it for every
task. They commonly skipped SDD for small tasks, bugs, and one-off scripts:
``If you're using [an SDD framework] for smaller changes, you're going to
have a bad time. It's too verbose and produces too much overhead and
documentation'' (R33). Among interview participants, tool users who described
how they decide mostly did so by \emph{task size or complexity}, using SDD
for large features. Urgency also overrode the workflow: for a production
hotfix, ``there is no time for spec-driven development and the whole flow''
(P4). Participants fixed bugs by prompting the agent directly and skipped SDD
for non-production work, such as one-off scripts and prototypes. Some tool
makers and some tool users decided to use SDD depending on \emph{risk or
uncertainty} associated with the task: P20 mentioned that the cost of failure determines how rigorous the spec must be. These participants advocated for SDD when the stakes associated with the task were high and the underlying intent was clear. 

\begin{figure*}[t]
  \centering
  \includegraphics[width=0.8\textwidth]{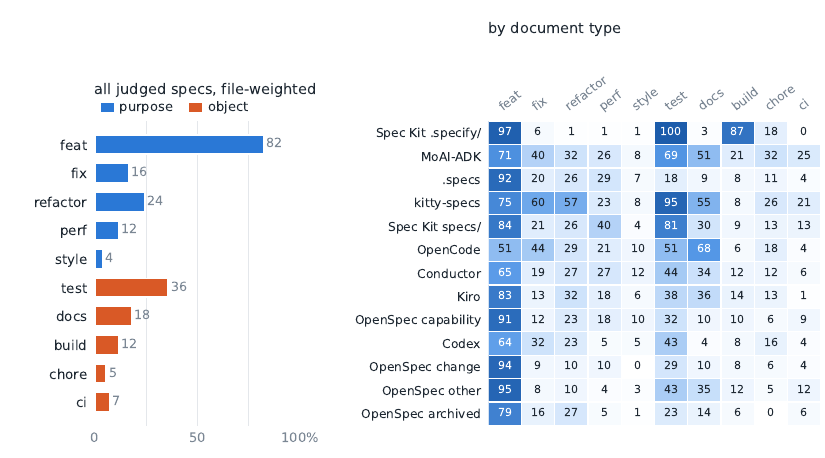}
  \caption{Purpose and object labels of 1{,}000 judged specs, following Watanabe et
  al.~\cite{watanabe2025agentic}; a spec may carry several labels. Left: shares weighted by each
  document type's number of files. Right: shares within each type.}
  \label{fig:tasktype}
\end{figure*}

The preference for specifying new features also appears in the repository
sample. We estimate that 82.0\% of specs describe new features, weighting each document type by
its number of files (Figure~\ref{fig:tasktype}), consistent with the qualitative findings (Appendix~\ref{app:specmine-results}).

\textit{Finding 3: Practitioners differ in how closely they read
specs, and most verify code mainly through agent-written tests.}\src{R, I, S}

\textit{Spec review.} Most tool users read specs only partially or
delegated their review to other agents, through adversarial review by other
models, prompting an LLM to find gaps, or asking an AI to check whether
concerns were addressed. On the other hand, some tool makers and a few tool users reviewed every spec in full detail. We observed that participants who reviewed specs in full typically worked in organizations where spec review was a team activity that involved coordination among product managers, senior engineers, and tech leads. Consequently, the differences in the extent of spec review may reflect the organizational setting as much as their individual roles as tool makers or users.

The interviewees situated spec review by multiple people into organizational
workflows. In the repository sample, the visible record centered on
individual authors and automated comments. We observed that 98.3\% of spec files with a recorded
history have a single author, and 68.5\% of inline comments on specs were
written by bots. These patterns show limited recorded human participation, but
not how closely people read specs or whether review happens elsewhere
(Appendix~\ref{app:specmine-results}; Section~\ref{sec:threats}).

\textit{Code verification.} Nearly all participants verified code through
agent-written tests, mostly end-to-end or acceptance tests. Only some participants also reviewed the resulting code by hand (Appendix~\ref{app:generic}). Many
found verification difficult or impossible: ``none of those tests are written
by hand'' (P3), and ``there's no way to, like, hold an AI accountable''
(P18). Participant P14 called this ``verification debt.'' Reddit posts likewise
urged checking agents' claims: ``Stop trusting what the agent says, verify it
with code'' (R85).

\textit{Finding 4: Practitioners differ in how they retain specs.
In public repositories, most spec files have a single recorded commit, and archived change records predominate in OpenSpec.}\src{R, I, S}

Participants describe one of the following three practices toward spec maintenance: (1)
Some maintained a \emph{living spec}, retained and updated as the system changed.
(2) Some kept specs in an \emph{accumulated change record}, committing a new spec to the record for
each change while retaining earlier specs. (3) A few \emph{discarded} specs after implementation and kept a
design record in an issue tracker, in folder
documentation, or in acceptance tests. Tool users were divided across the
three practices, whereas most tool makers kept living specs. Reddit discussions
showed a similar divide, and some posts argue for discarding specs because
``every SDD framework gets the same complaint: spec rot'' (R96). The tools also prescribe different practices. OpenSpec keeps both a living spec and an accumulated change record, Kiro keeps a living spec for each feature, and Spec~Kit documented the choice only in June 2026 (Appendix~\ref{app:tools}).

\begin{figure*}[t]
  \centering
  \includegraphics[width=0.8\textwidth]{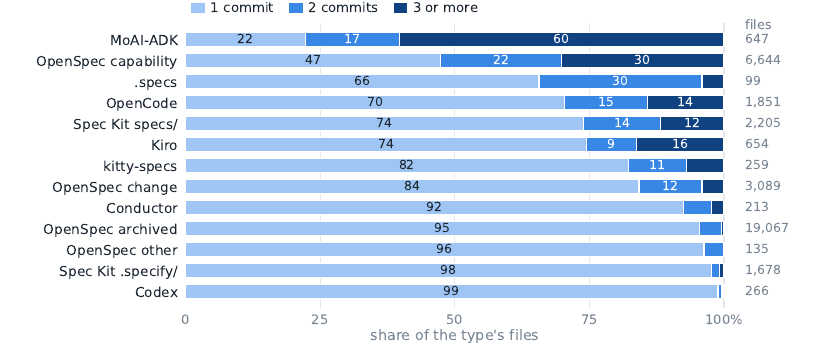}
  \caption{Recorded commits per spec file at its current path, by document type.}
  \label{fig:revision}
\end{figure*}

The distinction between living specs and accumulated records helps interpret
the files retained in the repository sample. 81.5\% of spec files have a single
recorded commit, and the share varies with where a tool files a spec
(Figure~\ref{fig:revision}; Appendix~\ref{app:specmine-results}). OpenSpec is the only tool whose
layout separates capability specs from archived changes,
and its repositories keep both. Archived changes outnumber capability specs
about three to one (19{,}067 vs.\ 6{,}644), because a project records each
change as a new file while revising the one capability spec it keeps.

\textit{Finding 5: Practitioners customize their workflows and
what specs cover.}\src{R, I, S}

Almost no participant used an SDD tool as published. Tool users extend
existing tools by adding their own skills and workflow layers, encoding
project conventions in configuration files such as
Kiro's steering files, or by adding their corporate framework on top. Workflows were customized to fit their organization's standards (``shared plugins inside my
organization's GitHub,'' P4) or their personal work style, and they
generally preferred lightweight, customizable tools. Tool makers instead built their
own tools because existing tools
did not fit their approach (``I tried it and I hated it, so I forked it,''
P11). ``Building one's own SDD tool'' was also the most frequent code that we assigned to
Reddit posts.

\textbf{Customizations based on the scope of individual specs.} We use \emph{feature} for a part of the system's behavior and \emph{change request} (or \emph{change}) for a unit of work, which can modify several features. Some participants' views regarding the scope of an individual spec file differ from the scope adopted by the two most widely used tools for SDD, namely OpenSpec and Spec~Kit. Both tools start a new spec for each change request. Spec~Kit creates a directory for each change request, which it calls a numbered feature. OpenSpec users write change specs for each change, which may modify several features, and the tool later merges them into the capability specs. Spec~Kit leaves to the team whether a spec stays in an accumulated change record or is updated as a living spec. Tool users
who kept to these tools scoped their specs to change requests (``we make, like, three
or four files per feature,'' P10). On the other hand, some participants who built their own
workflows, tool makers and some tool users, believed that a spec file should be scoped to features and not individual change requests and be updated as the system changes (P2, P3, P6). 
For several participants, how a tool records intent was a reason to customize: Participant P6 kept use cases as ``the source of truth,'' unlike workflows
that ``archive or delete the specs'', P14 moved the description of the system into a document in every folder
because specs ``are often likely to get out of date,'' and P12 kept
system-level design documents alongside specs.

\begin{figure*}[t]
  \centering
  \includegraphics[width=0.8\textwidth]{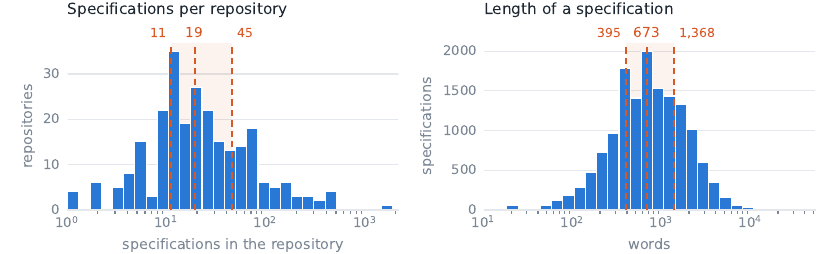}
  \caption{Specs per repository and spec length on log axes, as a
  proxy for spec scope.}
  \label{fig:scope}
\end{figure*}

In the repository sample, we conjecture that how a project organizes its spec files indicates whether the specs are scoped to change requests or the system's features. Our hypothesis is that a project that keeps one spec per feature should hold a few
long documents and one that writes a spec for each change request should hold many
short ones. Across 14{,}599 specs in 257 repositories, excluding OpenSpec's accumulating change records and its 135 files that match no lifecycle role, a repository holds a median of 19 specs of 673
words each (Figure~\ref{fig:scope}). The change-scoped spec that the most widely
used tools produce thus appears to be the common case in public repositories,
consistent with the interviews, although the proxy cannot separate a system
with many features from one with many changes
(Appendix~\ref{app:specmine-results}).

\subsection{RQ3: Benefits of SDD}\label{sec:findings-rq3}

\textit{Finding 6: Some participants reported improved shared understanding when specs passed between developers and other roles.}\src{I}

Collaboration benefits came from the few organizations where
specs pass between developers and team members employed in roles besides software development. A product owner told P6, ``Now I
see what my developers are doing;'' for P10's team, ``A lot of communication
has been offloaded to the spec itself;'' and P16's team depended less on
product managers, architects, and scrum masters as a consequence of adopting SDD. In developer-only teams, the benefit was modest:
specs became one more artifact to read during review, and some
saw little change (``Not a ton,'' P14). For solo developers, collaboration was with
the agent: ``Between me and the other humans, no. Between me and the coding
agents, yes'' (P17). Tool makers mostly described collaboration as one purpose
of specs, not a change that they had observed.

In the repository sample, repository owners and the three most frequent committers in each repository account for 90\% of the 13{,}640 spec commits. Recorded authorship is therefore concentrated. Commit records do not show the roles of the people who read or discuss a spec, however, so they cannot show how common the cross-role benefit is in open source (Appendix~\ref{app:specmine-results}).

\textit{Finding 7: Most practitioners found that specs made agent
  work more reliable, although they could not separate this benefit from
improvements in models.}\src{R, I}

Most participants described specs making agent work more reliable. Tool users
described it in terms of what the agent produces: better code with less
debugging and rework, and larger or longer tasks completed with less
supervision. Tool makers described what specs give the agent, such as
persistent context, so that the agent does not start ``from zero every time''
(P9), or orientation in a large codebase. A few did not report the benefit:
P18 found that ``direct prompting, I think, is way better'' than detailed
plans. Participants could not separate the benefit from concurrent
improvements in models; P8 concluded that ``this has largely to do with
models becoming so much better.'' Reddit discussions added more skepticism
than the interviews. After using SDD for months, one team concluded: ``We
couldn't prove it improved the code'' (R41); other posts found no clear gain
over direct prompting.

\textit{Finding 8: Drafting a spec with an agent helped some
practitioners clarify their intent, which kept developing as they built.}\src{R, I}

Because agents drafted most specs (Appendix~\ref{app:generic}), clarifying
intent happened in conversation with
the agent. The agent asked questions and produced a draft, and the
participant read and corrected it. Some participants found that this process
made their intent clearer: the ``Act of authoring a specification clarifies
the objective even to yourself'' (P9). Reading the agent's draft also showed
how the agent had understood them and what they had not yet decided:
P11 reads the spec asking ``is that what I want?'', and P17 has ``a
productive conversation with them about what we're trying to achieve by
forcing them to write a spec which I read.'' Intent did not become fully
clear before implementation. Participants refined specs over several
iterations and as implementation showed what was needed, and when they had
only a vague idea, some prototyped first: ``the more you create, the better
you'll become at knowing what you want to create'' (P2). A few found
articulating intent a burden: ``considering those considerations in your mind
is not enough'' (P10). On Reddit, practitioners likewise valued the agent's questions, as in a
workflow that ``asks good clarification questions'' (R89).

\subsection{RQ4: Challenges of SDD}\label{sec:findings-rq4}

\textit{Finding 9: SDD shifts token cost, time, and cognitive load
from implementation to planning.}\src{R, I}

Most participants discussed token cost, time, or cognitive load, and many
described a trade-off: SDD spends more on planning in exchange for less
rework and review later.
\begin{itemize}
  \item \textbf{Tokens.} One team comparing SDD with direct prompting found
    that ``the spec-driven runs typically consume two to three times as many
    tokens and take about twice as long'' (R41). Most tool users raised token costs, and the cost specific to SDD lay in
    planning: ``80\% of tokens are being used to make a plan, that is not okay''
    (P3). Reported savings were rework avoided, not a smaller bill (P13).
    Those paying for tokens themselves were most concerned.
  \item \textbf{Time.} Time moved from implementation to specification:
    ``Implementing the actual code is the easiest part right now'' (P10).
    Participants credited faster delivery to AI rather than SDD, and
    some doubted it because verifying code ``requires the same
    number of hours'' (P14).
  \item \textbf{Cognitive load.} Several tool users reported more load: long
    agent-written specs were hard to read and human attention was the
    bottleneck (P3). Some tool makers reported less, since reviewing a
    concise spec replaced code review, which P20 found ``cognitively
    expensive.''
\end{itemize}

\textit{Finding 10: Agents departed from specs during implementation and specs
  went stale as code changed; practitioners fixed whichever they blamed, their
instructions or their specs.}\src{R, I}

Most participants reported that code and spec diverged, in two ways. In
\emph{implementation drift}, reported mostly by tool users, the agent departs
from the spec while implementing it: it ``just goes off the rails and starts
doing all kinds of stuff that I didn't ask it to do'' (P5). In \emph{spec
decay}, the spec goes stale as the code changes: ``the code part will be
mutated, and the spec will be left alone'' (P2). If a spec says a
session expires after 30 minutes and the agent implements 60, that is
implementation drift; if a later change sets 60 as intended and the spec
still says 30, that is spec decay. Spec decay affects
living specs, mainly when changes alter behavior that other specs
describe, since a small change ``might be related to 10 other features''
(P10). Reddit posters who saw implementation drift blamed their
own instructions or the model. They responded to drift by making implicit
conventions explicit in instruction files. Participants who saw spec decay blamed incomplete or unlinked specs, and they
responded by limiting their specs or by binding specs together. Checks
such as scheduled agent reviews or end-to-end tests in continuous integration
can show that code and spec disagree, but not which is wrong.

\textit{Finding 11: The effort of evaluation grows with the system
and outgrows human ability to review.}\src{R, I}

Evaluation concerned three objects: the \textit{spec}, whose correctness
participants found hard to judge; the \textit{code}, which left verification debt (Finding~3); and
the \textit{consistency} between spec and code, which required the checks in
Finding~10. The volume of all three outgrew human ability to review. As a project grew,
``the documentation drifts, context balloons, and token costs climb'' (R55),
and agents produce ``a thousand lines of changes in one operation, I'm not
reading it'' (P11). In existing systems, a spec can become larger than the
code, and features that spanned several specs encountered bugs during implementation, unless
their order and dependencies were managed. 

\textit{Finding 12: Participants who introduced SDD to their teams as individuals often met inertia.}\src{I}

Participants who work in teams and organizational settings raised \textit{inertia}, which we define as the tendency to resist a change in their existing SE workflows, as a barrier to the adoption of SDD. However, inertia was not raised as a barrier by participants who were solo-developers or those who practiced SDD alone. We observed that individual adoption can arise before the team adopts SDD: ``For me, it was successful, because I enjoyed it,''
but ``you need the whole team to switch'' (P21). Most participants who reported inertia had introduced SDD to their team themselves. Some participants worked in organizations that adopted SDD as a process, by piloting it for a sprint, mandating it, or fitting a tool into an existing process. Inertia also appeared in these settings, so our data do not show that adopting SDD as a process avoids it. P12, whose team already followed a formal requirements specification process before implementation, did not report inertia. Tool makers and tool users described the inertia differently. Tool users named concrete barriers: setting up environments, working from the
command line, and anxiety about agents taking over parts of the job. Tool
makers described that a successful adoption of SDD requires a cultural shift in how team members collaborate with one another, which organizations mistake for a 
tooling choice (P9). A lack of
evidence reinforced the inertia: P6 wanted research to convince customers ``because people don't believe experience.''

\section{Discussion and Implications}\label{sec:discussion}

We discuss four themes that cut across the research questions: how much a
spec should say (Findings~1, 2, and~5), how intent is preserved as a system
evolves (Findings~4 and~5), how specs and code are kept consistent
(Findings~3, 10, and~11), and what SDD does for individuals and for teams
(Findings~6--9 and~12), each with implications for tool builders,
practitioners, and researchers. The interpretations rest on associations
across participants and repositories; our data do not establish their
direction, and the recommendations should be read with that limit in mind.

\subsection{How Much a Spec Should Say}\label{sec:disc-detail}

Consistent with the requirements engineering literature, tool makers keep \emph{what to
build} in the spec and \emph{how to build it} in a design or plan. Most tool
users put the \emph{how} in the spec, as do 77.6\% of the specs we judged
from the repository sample (Finding~1). Three explanations fit our data. (1)
Participants added implementation detail where they judged the agent incapable (P10) and where
agents lacked context that could not be inferred, such as internal APIs and domain rules
(P16). Implementation detail added because the agent performed poorly on a task may shrink as models improve. (2) Separation
may be unrealistic: requirements and architecture develop together rather
than in sequence~\cite{nuseibeh2001twinpeaks}, and participant intent continues
developing as software is built (Finding~8), so a spec written before design may be
unable to separate the what and how. (3) The tool
decides where the detail is located, since the share of specs with implementation
sections varies widely by tool (Finding~1).

Including \emph{how} in a spec also has a cost that no participant mentioned. We found that 75\% of the specs with implementation detail include named classes, functions or library versions, which change under
refactoring while the intended behavior does not. Thus, a spec that specifies
more implementation detail decays sooner (Finding~10). Participants added detail for complex tasks and skipped SDD for simple
tasks (Finding~2), so changes judged simple when they were made contained the least detail, although even a small change can affect many features (P10).
Each participant set their own threshold, no one described a team-level rule
for what warrants a spec, and a hotfix made without time for SDD 
changes behavior without a spec (P4). Sparse or missing
specs for such changes may add to what Storey calls \emph{intent
debt}~\cite{storey2026debt}.

\textit{Implications.} Tool builders should separate the parts of a spec by expected longevity: keep what-level content in a living spec,
mark implementation sections as scoped to the change, and make implementation sections optional
in templates. Practitioners should keep stable team knowledge, such as
conventions and internal APIs, in shared instruction files, as several
participants did in response to drift (Finding~10), record the rationale
behind consequential \emph{how}
choices~\cite{parnas1986rational,jansen2005decisions}, and agree as a team on
what warrants a spec, instead of leaving these decisions to each developer. Future research
should study how varying the amount of implementation detail in spec files impacts intent capture and later maintenance effort across model versions---keeping the task, model, and acceptance criteria fixed. Coding agents can also update project documentation and memory files such as \texttt{CLAUDE.md} as part of a change, so future research should examine whether specs replace, complement, or duplicate these artifacts.

\subsection{Preserving Intent as Systems Evolve}\label{sec:disc-record}

Recording design intent while building software serves two purposes:
(1) it describes what the system is currently meant to do; and
(2) it explains why the implementation decisions were made~\cite{parnas1986rational,jansen2005decisions}.
Parnas and Clements argue that design documentation should present
the system as if it had been designed rationally and kept current,
while still recording the alternatives considered and why they were
rejected~\cite{parnas1986rational}.
Most of the participants we interviewed mentioned that preserving
design intent in spec files is a key benefit of SDD (Finding 8)
and highlighted two primary ways of maintaining specs over time:
as a living spec and as an accumulated change record (Finding 4).
These approaches serve different purposes.
A living spec describes the current intended behavior of the system,
whereas an accumulated change record preserves how the design evolved
over time. The former is closer to rational design documentation,
while the latter is closer to the actual project history.

Current tools for SDD differ in how they maintain these records.
OpenSpec keeps both the living spec and the accumulated change record
as separate files in the working tree. Kiro updates one spec per feature as
requirements change, so it keeps a living spec and leaves the accumulated
change record to version control. Spec~Kit leaves the choice between the two to the team. However, recovering design history from version
control can be lossy when files are renamed or commits are squashed
(Section~\ref{sec:threats}).
Prior research in human-computer interaction has established that an artifact's
design features influence which actions users perceive and can
readily enact, even when the artifact supports a broader range of
actions~\cite{norman2013design}.
Similarly, how specs are maintained in repositories may influence
whether developers consult the current design or revisit earlier
decisions. Keeping accumulated change records in the working tree
makes design history more accessible, but also introduces a tradeoff
for AI coding agents, whose context typically holds only files in
the working tree. As archived records grow with every change,
agents may encounter outdated intent that is no longer relevant
to the current implementation (R55).
Keeping change records in version control avoids introducing
historical information into the working tree, but makes design
history less accessible, particularly because reviewing prior
commits is not common practice among developers.

A further limitation concerns the information preserved in these
records. Existing tools and practices cannot recover design
rationale that was never documented. Moreover, most archived
change records are agent-drafted descriptions of what changed,
rather than explanations of why the change was made. Consequently,
preserving a history of changes does not necessarily preserve the
reasoning behind design decisions.

\textit{Implications.}
Tool builders should distinguish between maintaining current design
intent and preserving design history. Living specs should remain
up to date and readily accessible to coding agents, while accumulated
change records should remain available for developers to inspect
when investigating earlier decisions. Tool designs should balance 
the accessibility of design history against the risk 
of exposing agents to outdated intent. 
Furthermore, tools should also encourage developers
to document the reasons behind significant changes.
For practitioners, maintaining useful design documentation requires
updating living specs as the system evolves and recording the
rationale behind important design decisions, rather than relying
solely on automatically generated change descriptions.
Future research should investigate how different approaches to
storing and presenting design history affect developers' ability
to recover earlier decisions and agents' ability to use current
design intent without being distracted by outdated information.

\subsection{Keeping Specs and Code Consistent}\label{sec:disc-consistency}

A living spec is useful only while spec and code agree, and participants reported
two ways they diverge: the agent departs from the spec while implementing it,
and the spec goes stale as later changes alter the code (Finding~10). The tools
in our repository sample address implementation drift. For example, OpenSpec merges the specs of a change into the capability specs before archiving
the change and provides a verify command that checks the implementation against the change, Kiro
gave one participant requirements traceability, and two tool makers bound
specs to code through hashed contracts (P2) or a curating agent (P11). Our study did not establish whether any of these tools checks that specs and code still agree after the implementation is complete. Participants who wanted
to detect later decay built their own checks (Finding~10), which were costly
and, when agents ran them, unreliable, as when one verify command worked about half the time (P20).

We doubt that current approaches scale, for three reasons. First, spec decay
can possibly affect the specs other than the spec being implemented, and finding these other specs is a traceability problem~\cite{gotel1994traceability} compounded by SDD that produces more specs faster. Second, the spec corpus grows like a codebase (Finding~11), and a check that compares natural-language specs against code must interpret both spec and code. An agent may find that the code is consistent with an ambiguous requirement because the ambiguity admits multiple implementations. Third, inconsistencies between specs and implementation may produce few observable signals. For example, an agent modifying a feature may introduce behavior that satisfies the immediate task and passes its associated tests but violates a constraint documented in an earlier spec. Such inconsistencies may remain undetected despite running tests because tests typically verify specific expected behaviors rather than all the goals, constraints, and rationale captured in specs. Furthermore, even consistency checks may only identify disagreements between specs and code and not determine which of the two should change, leaving the decision to a person. This reflects Bainbridge's irony of automation: humans remain responsible for monitoring work in which they are increasingly less involved~\cite{bainbridge1983ironies}.

The human reviewer must also read the spec, yet most tool users read specs partially or delegated review to agents (Finding~3). Partial reading is partly what SDD intends, since parts of a spec carry different weight (P20): the problem statement, key decisions, and acceptance criteria state intent, while the rest is agent context. Most participants who set or checked acceptance criteria, or reviewed the agent's tests, did not raise verification as a challenge. We cannot show these practices prevented defects, but the association suggests that human effort may be better spent on setting acceptance criteria that define correct behavior than on inspecting code line by line. Partial reading is also a risk: when one agent writes the spec, the code, and the tests, one misunderstanding can propagate to all three, so some participants insisted that an agent must not review its own implementation (P21). Delegating spec review to an agent removes the one point where a person decides what the system should do. Over-reliance on AI falls when a person must act before accepting its output~\cite{bucinca2021trust} and reading the spec is that act in SDD.

\textit{Implications.}
To reduce implementation drift, tools should turn acceptance criteria into tests the agent must pass and link spec clauses to tests and code for traceability. To reduce spec decay, tools should detect code changes linked to specs, archived specs, or changes made without specs, then draft updates from tickets, prompts, and diffs for human confirmation. Tools should also separate specification, implementation, and review contexts, present what-level sections first, and show acceptance-criteria revisions as diffs. Practitioners should write or approve acceptance criteria even when agents produce the rest. Future research should measure drift in SpecMine by examining pull requests that modify both specs and code and OpenSpec records that identify modified capabilities. It should also compare human-written, human-reviewed, and unreviewed agent-generated acceptance criteria based on the defects their tests miss.

\subsection{Benefits for Individuals and Teams}\label{sec:disc-team}

Most participants reported more reliable agent work after specs, but they could not
separate that benefit from model improvement (Finding~7). A team that
measured SDD could not show that SDD improved the code (R41). One explanation
consistent with our data is that the benefit relative to vibe coding comes
less from the agent having a spec than from the developer having articulated
goals before delegation (Finding~8). SDD would then work as cognitive forcing~\cite{bucinca2021trust}, and the same benefit should appear with any
practice that leads a developer to plan before acting, including plan
modes present in AI coding agents like Cursor and Claude Code. Participants valued when agents asked clarifying questions (Finding~8), yet no participant described tooling that presents design alternatives or records rejected designs.

Tool makers emphasized collaboration as a purpose of specs more frequently than tool users reported using specs for collaboration. Some participants reported increased shared understanding when specs passed between developers and other roles (Finding~6). Accountability and trust from shared specs also depend on teammates reading specs and agents following them, which may be difficult given partial reading and drift (Findings~3 and~10). The work and benefits also fall on different people~\cite{grudin1988cscw}. The author writes the spec, while later teammates or agents receive the benefit. This may explain why individuals persisted with SDD while their teams resisted it, because the extra work is visible and the shared benefit is not. Participants who introduced SDD to their teams as individuals often met inertia (Finding~12), consistent with diffusion research on innovations whose relative advantage is hard to observe~\cite{rogers2003diffusion}. Individual success was invisible to the team (P21), and others discounted experience as evidence (P6). Organizations that adopted SDD as a process piloted tools for a sprint or integrated them into existing processes. Whoever pays for tokens bears the planning cost, while later maintainers gain from avoided re-work (Finding~9).

\textit{Implications.} If articulating goals produces individual
benefit, then tool builders should support drafting these goals. The
agent should elicit design alternatives and trade-offs by default and retain context behind
the rejected designs, which comprises the otherwise undocumented rationale (see
Section~\ref{sec:disc-record}). Organizations
should adopt SDD as a process by introducing roles expected to read specs,
piloting SDD within existing processes, and evaluating communication
across roles alongside delivery time, review effort, token cost, and rework. Future research should test the
articulation explanation by comparing a spec written and provided to the agent,
a spec written and withheld, and no spec, on the same tasks across model
versions, and follow teams over time to see which handoffs specs improve and
whether benefits persist. We release SpecMine to support this aim.
\vspace{-2mm}

\section{Threats to Validity}\label{sec:threats}

\subsection{Internal Validity}

We identify associations, not causal effects. Tool makers and tool users
differ in experience, roles, and organizational settings, and these
differences may shape their SDD practices. Reported improvements also
coincided with improvements in model capability, so we cannot attribute them
to SDD alone. We therefore treat explanations of benefits and difficulties as
interpretations of practitioner accounts.

Our judgment as researchers may have influenced the interviews and the
coding. Each source was first coded by one author, and we did not measure
inter-rater agreement (Section~\ref{sec:method-interviews}). Merging the
codebooks in discussion among all authors and re-coding all data against the
merged codebook reduces this influence but does not remove it. The
interviewer did not introduce a theme before a participant raised it. We
checked the LLM-corrected transcripts against the originals, but this check
cannot rule out every change in meaning.
\vspace{-2mm}
\subsection{Construct Validity}

Reddit discussions and interviews capture reported experiences, which may
differ from actual practice. The repository sample does not measure benefits,
effort, or perceived challenges, so the three sources can be compared only on
claims that repositories make observable. The sources also differ in
population. Interviewees are practitioners engaged enough with SDD to
volunteer, and the repository sample holds public, popular, OpenSpec-heavy
repositories. Agreement between the sources therefore corroborates a finding,
but a disagreement may reflect a difference in either practice or population.
Shares of files are not shares of practitioners, and a single repository may
hold many archived changes alongside a few living specs.

The repository measures capture the findings only indirectly. Scope is
identifiable only where a tool's layout encodes it, as OpenSpec's separate folders for capability specs and change records do. Headings show how specs
are structured, not what they say or how good they are. The content and
purpose labels come from an LLM and may be wrong. Review is visible only as
inline comments on GitHub, so a spec with one author or few comments may
still have been discussed elsewhere.

Commit counts understate revision. The histories we retrieved do not follow
renames, so a file moved to a new path loses the commits made before the
move. OpenSpec's archive step moves every completed change in this way.
Squashed or rebased commits also merge several edits into one. Conversely, a
recorded revision may only fix a typo. Commit counts therefore cannot show
spec decay, which would require checking whether the behavior a spec
describes changed while the spec did not.
\vspace{-2mm}

\subsection{External Validity}
We recruited interview participants through professional networks, online
communities, and projects that use specs, so they are likely more engaged
with SDD than practitioners generally. Their accounts show which practices
exist but not how common the practices are. Our choice of Reddit communities,
keywords, and popular records may favor visible experiences over less common
perspectives. Our Reddit sample is purposive rather than exhaustive. Reddit's search
ranking is undocumented and favors recent and highly engaged content, so we
cannot quantify what it did not surface, and we report no community totals
because the interface exposes none. Deleted and removed content is invisible
to search, and a record no query surfaced is absent without trace.

SpecMine covers public GitHub files that code search can find. Deleted specs
are absent, which particularly limits comparison with the temporary specs
that interviewees described. The repository sample further selects
repositories by popularity, license, spec activity, and recognized tool
layout. Its results may not represent private projects, brief experiments,
or other layouts. Comparisons between capability specs and archived changes
rely on OpenSpec, which accounts for 78.4\% of the sample, and may not extend
to other tools.

Document types also differ sharply in how many repositories they draw on.
OpenSpec's three roles and Spec~Kit's numbered features each span 67 to 155
repositories. In contrast, six of the 13 types take at least half their files
from one repository. For OpenCode, Codex, and Spec~Kit's \texttt{.specify/}
scaffold, that single repository supplies more than 95\% of the files, and
two of these concentrated types come from the tool vendor's own repository.
Results for those types describe how one or two projects apply a template,
not a practice observed across many projects.

The study covers an early period of SDD adoption. Interviews took place in
2026, and the repository histories span less than a year of adoption. Tools,
models, and organizational practices changed during the study, so the
findings characterize this period.
\vspace{-2mm}

\section{Conclusion and Future Work}

Spec-driven development has spread quickly, but evidence about how it is
practiced comes mostly from the vendors who promote it. Through an
analysis of Reddit discussions, interviews with 21 practitioners, and 266
active and popular repositories in SpecMine, a corpus of
470,795 spec files that we release, we characterized spec content, workflows, benefits, and challenges.
Tool documentation promises better quality, less agent drift, and higher
productivity. Participants reported more reliable agent work but could not
separate this benefit from model improvements (Finding~7), agents still
departed from specs (Finding~10), and SDD shifted cost to planning rather than
removing it (Finding~9). The difficulties that limited earlier
spec-first practice also persist: specs cost effort to write, go stale
as code changes, and remain hard to trace to the code that satisfies them. Practitioners used specs selectively and differed in
how they maintained them, relied on custom checks to detect divergence after
implementation, and reported collaboration benefits mainly where a team adopted
SDD together. Future work should examine how practitioners identify necessary
spec updates, whether human involvement in acceptance criteria improves
verification, and how the benefits of SDD change as models improve and teams
adopt it. We release SpecMine to support this work.

\section{Data Availability}
A replication package is available at \url{https://github.com/shyamagarwal13/spec-paper-replication-package}.

\section{Acknowledgments}

This work was supported in part by NSF Award \#2217572, CMU S3D Presidential Fellowship, and Google research gift.

\bibliographystyle{IEEEtran}
\bibliography{references,specmine-refs}

\onecolumn
\appendices

\begingroup
\setlist{nosep,leftmargin=1.4em}

\section{The \dsname{} Corpus}\label{app:specmine}
\noindent\textit{\emph{Spec-Driven Development} (SDD) is a fast-emerging practice in which a
structured natural-language specification, written by a developer, or (more
often) drafted by an AI tool and then curated by the developer, drives an AI
coding agent's implementation. A wave of tooling (GitHub Spec
Kit~\cite{speckit}, OpenSpec~\cite{openspec}, AWS Kiro~\cite{kiro},
and dozens of others) has appeared since 2025, yet the artifacts
these tools produce have never been studied at scale. We present \dsname{}, a
corpus that captures SDD in public GitHub repositories through two censuses: a broad census of
\texttt{spec.md}/\texttt{specs.md} files covering most tools
(\num{470{,}795} files across \num{73{,}030} repositories, attributed to 17 named tools), and a Kiro census of its distinct
\texttt{requirements}/\texttt{design}/\texttt{tasks} layout (\num{98{,}574}
files across \num{12{,}910} repositories). Each spec in the broad census is
enriched with full repository metadata, complete commit history, and parsed
document structure.
How a spec becomes code is itself an open question, so for 11 tools we sweep
every pull request that touches a spec in their repositories with at least ten
stars, capturing \num{5{,}979} such PRs across 581 repositories with their
changesets. The sweep makes the simplest workflow directly observable: spec and
implementation changing together in one PR. A census-wide index of
\num{2{,}421{,}323} typed references gives a second, independent link from spec
to code (1.28M to code files, 863k to sibling documents, 152k to PRs, 62k refs,
43k branches, 22k issues).
\dsname{} lets the community study, for the first time, how software is
specified in the age of AI agents.}

\subsection{High-Level Overview}

\subsubsection{Why spec-driven development, and why now}
Spec-Driven Development (SDD) tools help a developer capture what to build as
one or more Markdown documents that an AI agent then uses to generate and modify
code. Some tools keep this in a single specification file; others, such as Spec
Kit and Kiro, split it into separate requirements, design, and task-list documents. 
The spec itself can be human-written, but is often augmented or generated by AI
from a short prompt, brief, or product requirements document, then modified by a
human.
The practice barely existed before 2025; it now spans more than 70,000 GitHub
repositories and dozens of competing tools. The spec, rather than the code, is
becoming the primary artifact developers write, review, and maintain. Yet
existing datasets do not capture these artifacts, identify the tool that
produced them, or show how they relate to the code that ships.

Two MSR Challenges have looked at neighbouring artifacts of AI-assisted
development: DevGPT (MSR~2024)~\cite{devgpt} at developer--ChatGPT
\emph{conversations} and AIDev (MSR~2026)~\cite{aidev} at
\emph{agent-authored} pull requests. But both study the
model's output. In this paper, we present \dsname{}, the missing \emph{intent}
layer: the specification of
\emph{what} to build (whether a developer writes it or drafts it with an AI tool
and refines it), captured across 18 SDD tools (17 named plus Kiro), and, on a
subsample of mature repositories, the pull requests that touch it.

\subsubsection{What the dataset contains}
We capture SDD artifacts from public GitHub repositories through the code-search
and REST APIs, in four parts:

\begin{enumerate}
\item \textbf{The broad census.} Nearly every SDD
  kit stores its specification as a \texttt{spec.md}/\texttt{specs.md} file, so a
  census of that filename recovers the bulk of the ecosystem. Each kept file
  records the full GitHub repository object (stars, license, language, topics,
  timestamps), file/commit provenance, and an attribution to the
  tool that produced it (17 named tools such as OpenSpec, Spec Kit, and
  Conductor), assigned by a path fingerprint, the directory layout each kit
  generates. We manually checked 30 specs spanning ten named tools
  and found the attribution correct in every case: each spec is in its tool's
  directory and matches that tool's published template.
\item \textbf{The Kiro census.} AWS Kiro is among the most adopted SDD tools but is invisible to a
  \texttt{spec.md} search: it stores \texttt{requirements.md},
  \texttt{design.md}, and \texttt{tasks.md} under \texttt{.kiro/specs/}. We
  census that layout separately, recovering \num{98{,}574} artifacts across
  \num{12{,}910} repositories.
\item \textbf{The pull requests that change specs.} Across eleven of the named
  tools (\hyperref[app:qc]{Appendix~B} says which, and why those eleven), we sweep every
  repository with at least ten stars (949 repositories, 963 repository--tool
  targets, since a repository can adopt more than one tool) and capture every pull
  request that touches a spec file, together with its changeset, each file
  flagged spec or code, so a single PR shows whatever code changed alongside the
  spec. This yields \num{5{,}979} spec-touching pull requests across 581 of those
  repositories. We exclude the caffeine.ai app-generator (auto-generated single
  apps) and unattributed ad-hoc specs, neither of which reflects a deliberate SDD
  process. Because
  every row records live GitHub identifiers, the same procedure extends to more
  repositories, which we intend to do in later snapshots. How specs actually
  become code is an open question (\hyperref[app:qc]{Appendix~B}).
\item \textbf{The traceability index.} A spec's text, its commit messages, its
  OpenSpec \texttt{tasks.md}, and its folder's git tree all point to other
  artifacts: pull requests, issues, and the files the spec should change. We
  assemble these pointers into a census-wide index of
  \num{2{,}421{,}323} references, each typed by relation and provenance and
  rolled up onto the spec it came from. For OpenSpec, whose \texttt{tasks.md}
  names code paths directly, we also resolve every task-to-code
  reference against the repository's git tree at the anchoring commit
  (\num{435{,}401} references, each recording whether its task was checked off
  and whether the named file existed at that commit). These references are a
  second channel from spec to code, independent of co-change. A spec's commit
  that references PR \#123 points at an implementation even if that PR does not
  touch any spec file. A task naming a file that never appeared is evidence of
  a spec never implemented.
\end{enumerate}

Every spec in the broad census is further enriched with its full commit history
and 39 parsed
structural features (heading tree, code/table/diagram counts, requirement-%
template markers such as EARS~\cite{mavin2009ears} and Gherkin, and quality signals such as unfilled
placeholders and \texttt{TODO}s). The Kiro census ships as metadata in v1.0;
its commit history and features are in the 500-repository sample, not the full
release (Sec.~\ref{sm:sec:availability}). Attribution and enrichment are reversible and
idempotent.

\subsubsection{Summary statistics}
This paper describes \dsname{}~v1.0, the July~2026 snapshot. Nearly all
specs (99.7\%) were first committed in 2025 or later, matching when these tools
emerged, and 92\% in 2026, so the corpus captures the practice from its birth. At a glance it offers \textbf{scale} (\num{470{,}795}
specifications), \textbf{breadth} (\num{73{,}030} repositories across 17 named tools,
plus \num{12{,}910} Kiro repositories), and \textbf{depth} (every spec
with full commit history and 39 structural features, \num{266{,}230}
OpenSpec change artifacts, and, for 581 repositories, a curated layer
of \num{5{,}979} spec-touching pull requests with their changesets).
caffeine.ai has the most repositories but is an app \emph{generator} whose
repositories are largely auto-generated single apps; among developer-adopted
tools the leaders are Kiro, Spec Kit, and OpenSpec.
Table~\ref{sm:tab:overview} gives summary counts; Table~\ref{sm:tab:fulltools}
(\hyperref[app:tools]{Appendix~E}) gives the full per-tool breakdown of all 17 named tools, the 4 path
buckets, and the Kiro census.

\begin{table}[t]
\centering
\caption{Summary statistics.}
\label{sm:tab:overview}
\small
\begin{tabular}{@{}lr@{}}
\toprule
Quantity & Count \\
\midrule
Spec files (broad census) & \num{470{,}795} \\
\quad distinct repositories & \num{73{,}030} \\
\quad distinct owners & \num{44{,}521} \\
Named SDD tools (broad census) & 17 \\
Kiro artifacts (separate census) & \num{98{,}574} \\
\quad distinct repositories & \num{12{,}910} \\
Spec-file commits & \num{780{,}335} \\
Files with content + structural features & \num{468{,}307} \\
Spec-touching pull requests (11 tools swept, subsample) & \num{5{,}979} \\
\quad repositories with $\ge$1 such PR & 581 \\
\quad repositories swept ($\ge$10 stars, 11 tools) & 949 \\
\quad that also modify code in the same PR & 81.2\% \\
\quad per-file diff rows captured & \num{346{,}991} \\
OpenSpec change artifacts (proposal/design/tasks) & \num{266{,}230} \\
Typed references (traceability index) & \num{2{,}421{,}323} \\
\quad OpenSpec task$\to$code refs resolved on the tree & \num{435{,}401} \\
Licensed repositories & \num{28{,}698} \\
Repositories with $\ge$100 stars & 923 \\
Release size (download / uncompressed) & 5.4 / 14.1\,GB \\
\bottomrule
\end{tabular}
\end{table}

\subsection{Internal Structure}
The data are stored in a relational (MySQL) schema and are also released as flat
CSV/Parquet exports plus a JSONL of spec contents. The broad census is
the spine (\texttt{spec\_files}, keyed on a 16-hex digest of the file's GitHub
blob URL, \texttt{file\_url\_sha16}); the Kiro census and the PR layer attach to
it by \texttt{repo\_name}.

The remaining tables attach to this spine: \texttt{spec\_file\_commits} (per-file
commit history) and \texttt{spec\_content\_features} (the 39 features) keyed on
\texttt{file\_url\_sha16}; \texttt{kiro\_files} (the Kiro census);
\texttt{pull\_requests} and \texttt{pr\_files}
(the PR layer, built as eleven per-tool tables and released as two, each
carrying a \texttt{tool} column, keyed on \texttt{repo\_name}, flagging each
changed file \texttt{is\_spec}/\texttt{is\_code}); \texttt{openspec\_artifact\_files} and
\texttt{openspec\_code\_refs} (OpenSpec change artifacts and their resolved
task-to-code references); and \texttt{spec\_links} with \texttt{repo\_trees} (the
traceability index). Fig.~\ref{sm:fig:er} is the entity-relationship diagram and
Table~\ref{sm:tab:dict} a condensed data dictionary; \hyperref[app:schema]{Appendix~A} describes every
released table, and a data dictionary of the released tables (with types and
provenance notes) ships as \texttt{DATA\_DICTIONARY.md}.

\hyperref[app:qc]{Appendix~B} describes how we build both censuses, by adaptive
size partitioning around GitHub's 1{,}000-result cap, and how we check the
result for idempotence and determinism. It also documents the
\emph{not-a-spec} filter and its funnel (\num{822{,}901}
rows $\to$ \num{575{,}633} distinct $\to$ \num{470{,}795} kept), the composition
of the filename census and how to draw a maturity line through it, the scope and
sampling of the PR layer (a subsample of repositories), and the co-change heuristic
that treats a spec-and-code PR as implementing that spec, together with its
limits.

\subsection{How to Access}
\begin{itemize}
\RaggedRight
\item \textbf{Obtain.} Each snapshot (from v1.0) ships on Zenodo with a citable
  DOI as a MySQL dump, CSV/Parquet exports, and a JSONL of spec contents; a
  GitHub mirror has the schema, loader scripts, an example Jupyter/Colab
  notebook, and a 500-repository sample.
\item \textbf{Recommended use.} Load the dump into MySQL/DuckDB, or read the
  Parquet directly with \texttt{pandas}/\texttt{polars}: for instance
  \emph{``which specs change in the same PR as code?''} is a single join between
  \texttt{pull\_requests} (\texttt{touches\_code}=1) and
  \texttt{pr\_files}. Because every identifier is a live GitHub URL, participants
  can \emph{bring their own data}, re-fetching any row or joining external sources
  (issues, CI logs, registries).
\item \textbf{Skills.} SQL and basic \texttt{pandas} suffice, with no special
  hardware for the Parquet layer; a free GitHub token is needed only to re-fetch
  or grow the corpus. The download is 5.4\,GB and expands to 14.1\,GB; loading
  the dump into MySQL needs about 16\,GB of disk, the Parquet layer alone 1.6\,GB.
\end{itemize}

\subsubsection{Licensing, ethics, and privacy}
\label{sm:sec:ethics}
All artifacts come from public GitHub repositories, collected within the API
terms of service. Every row keeps its repository's license (\num{28{,}698} have
a recognized SPDX license), so participants can filter to license-compatible
subsets; redistribution follows each repository's terms.

The release ships public handles but not private contact details. Author and
committer logins and numeric IDs are kept: they are public, a repository's slug
already contains its owner's handle, and they are what makes authorship analysis possible
(bot detection, newcomer-versus-maintainer, the developer--agent division of
labour). Commit emails, avatar URLs, and the remaining profile fields are dropped
before release; a participant who needs them can fetch them from GitHub directly,
under GitHub's terms and their own institution's approval. Commit
messages and spec text are authored prose that may contain people's names, and
we do not rewrite them. We ask participants to avoid deanonymization and to keep personal
data out of published results beyond aggregate analysis.

\subsection{Research Questions}
\label{sm:sec:rq}
Making the specification a first-class unit of study, alongside the tool that
produced it and the code that implements it, opens agendas that a code-side dataset
cannot reach. We group representative questions below, each tied to the layer
that enables it. Each question is answerable empirically, and most cut
across several tool families.

\begin{enumerate}[leftmargin=1.35em]
\item \textbf{Adoption and diffusion of spec-driven development.}
  \emph{Enabled by the cross-tool census (18 SDD tools) over the
  2024--2026 emergence window.}
  \begin{itemize}[leftmargin=1.1em]
  \item Who adopts SDD (newcomers or experienced maintainers, which languages,
    domains, and team sizes), and how does adoption diffuse across ecosystems?
  \item How do the competing tool families grow, coexist, or displace one
    another, and do their document templates converge toward a shared structure?
  \end{itemize}

\item \textbf{Anatomy and quality of specifications.}
  \emph{Enabled by raw content plus 39 structural features across
  \num{468{,}307} specs.}
  \begin{itemize}[leftmargin=1.1em]
  \item What forms do specs take (EARS, Gherkin, user stories, free prose), and
    can a measurable notion of spec quality (completeness, testability,
    ambiguity, unfilled placeholders) be defined and validated?
  \item How much of a spec is reused template boilerplate versus
    project-specific content, and does that ratio differ by tool or author?
  \end{itemize}

\item \textbf{The spec--code relationship.}
  \emph{Enabled by two independent channels: the PR layer (\num{5{,}979}
  spec-touching PRs across 581 repositories, with complete, spec/code-flagged
  changesets) and the traceability index (\num{2{,}421{,}323} references from spec text, commit
  messages, \texttt{tasks.md}, and folder trees, of which we resolved
  \num{435{,}401} task-to-code references against the repository tree).}
  \begin{itemize}[leftmargin=1.1em]
  \item \emph{Where does the implementation of a spec live?} Co-change in a
    single PR is the obvious first guess and is common in the PR layer, but other
    workflows are equally plausible: a spec merged on its own and implemented in
    one or several later PRs, an implementation that arrives before the spec is
    written down, or a spec that is never implemented at all. Can these patterns
    be recognized and told apart at scale, how prevalent is each, and does the
    mix differ by tool, team size, or repository maturity? This is the question
    behind the heuristic of \hyperref[app:qc]{Appendix~B}, and we regard it as open.
  \item How often is a spec kept in sync as the code evolves, and when the two
    drift apart, which side changes first?
  \item A spec names the files it expects to change. How often does that code
    actually exist? Resolving \num{435{,}401} OpenSpec task-to-code references
    against the tree at the anchoring commit makes the gap between what a spec
    declares and what the repository contains measurable: what predicts it, and
    does it widen as specs get longer?
  \item Does the presence or quality of a spec correlate with PR size, review
    effort, or merge outcome?
  \end{itemize}

\item \textbf{Human--AI collaboration around specs.}
  \emph{Enabled by authorship, commit cadence, and template signals in the age
  of coding agents.}
  \begin{itemize}[leftmargin=1.1em]
  \item Can human-authored and agent-generated specs be distinguished, and how
    is authorship of a single spec shared between developer and agent?
  \item Do higher-quality specs predict smoother downstream implementation
    (fewer follow-up fixes), and how does review differ when the artifact under
    discussion is a spec rather than code?
  \end{itemize}

\item \textbf{Lifecycle, evolution, and abandonment.}
  \emph{Enabled by \num{780{,}335} commits of full per-spec history and
  lifecycle roles (living / proposed / archived).}
  \begin{itemize}[leftmargin=1.1em]
  \item What are the churn and half-life of a spec, and how often are specs
    abandoned mid-flight (open tasks that never close, placeholders never
    filled), and what predicts abandonment?
  \item Because every prior state is reconstructable from commit history,
    can spec-driven workflows be studied longitudinally without the data-leakage
    pitfalls of snapshot datasets?
  \end{itemize}
\end{enumerate}

\subsection{Data Availability}
\label{sm:sec:availability}
\dsname{} is one of the two datasets of the MSR~2027 Mining Challenge. Version~1.0 (the
July~2026 snapshot) is released in three tiers, so a user can start at whatever
scale suits the study:

\begin{itemize}
\RaggedRight
\item \textbf{Full dataset (Zenodo).} The complete corpus is archived on Zenodo
  with a citable DOI (\href{https://doi.org/10.5281/zenodo.22102779}{\texttt{10.5281/zenodo.22102779}}) as a MySQL dump plus flat
  CSV/Parquet exports and a JSONL of spec contents.
\item \textbf{Parquet mirror (Hugging Face).} A per-table Parquet mirror of every
  released table is published at \url{https://huggingface.co/datasets/ShyAgarwal/specmine}
  for direct use with \texttt{pandas}/\texttt{polars}/DuckDB, with no need to load
  the dump.
\item \textbf{GitHub mirror and sample.}
  \url{https://github.com/shyamagarwal13/specmine-official} has the schema
  (\texttt{schema.sql}), loader scripts, the complete data dictionary
  (\texttt{DATA\_DICTIONARY.md}), an example Jupyter/Colab notebook, and a curated
  500-repository sample. The sample is a faithful, reproducibly selected slice of
  every layer (\num{28{,}583} specs, \num{18{,}585} Kiro artifacts, \num{5{,}335}
  spec-touching pull requests with their \num{288{,}267} per-file changesets, and
  \num{261{,}032} traceability references), small enough to clone and query in
  minutes. It also carries the Kiro commit histories and features absent from v1.0.
\end{itemize}

Every identifier in the release is a live GitHub value (repository slug, blob
URL, pull-request number), so any row can be re-fetched or joined to external
sources. All artifacts are drawn from public repositories, each with its own
license (Sec.~\ref{sm:sec:ethics}); the \dsname{} compilation is released under
CC~BY~4.0. Later snapshots will be published as new versioned Zenodo records,
each with its own DOI, so results stay reproducible against a fixed version.

\phantomsection\label{sm:app:schema}
\subsection*{Appendix A: Schema and Entity--Relationship Diagram}
Figure~\ref{sm:fig:er} gives the entity--relationship diagram of the released
schema; Table~\ref{sm:tab:dict} is a condensed data dictionary of its core tables.
A data dictionary of the released tables ships as
\texttt{DATA\_DICTIONARY.md}. \hyperref[app:qc]{Appendix~B} documents how the corpus is
constructed and quality-controlled; \hyperref[app:vocab]{Appendices~C--E} give the data-specific detail
a participant needs to query the release directly.

The released tables, in full:
\begin{itemize}
\RaggedRight
\item \texttt{spec\_files}: one row per spec file: the full repository object,
  file/commit provenance, tool attribution (\texttt{spec\_tool}/%
  \texttt{spec\_role}), and raw content.
\item \texttt{spec\_file\_commits}: the complete per-file commit history
  (author, committer, timestamps, message).
\item \texttt{spec\_content\_features}: the 39 structural and
  requirement-template features per file.
\item \texttt{kiro\_files}: the Kiro \texttt{requirements}/\texttt{design}/%
  \texttt{tasks} artifacts, with their kind and feature.
\item \texttt{pull\_requests} and \texttt{pr\_files}: across the 11 swept tools
  (10 of which yield at least one PR), the
  spec-touching pull requests (state, merge, branch, line counts, and a
  \texttt{touches\_code} flag) and their per-file changesets (each file flagged
  \texttt{is\_spec}/\texttt{is\_code}). Both carry a \texttt{tool} column: the
  crawl wrote one table per tool and the release concatenates them.
\item \texttt{openspec\_artifact\_files} and \texttt{openspec\_code\_refs}: for
  the OpenSpec ecosystem, the full change proposals, designs, and task lists with
  parsed task counts, plus the task-to-code references they contain, resolved
  against the repository tree (\texttt{task\_done},
  \texttt{exists\_at\_anchor}).
\item \texttt{spec\_links}: the census-wide index of typed references, assembled
  from spec text, commit messages, OpenSpec \texttt{tasks.md}, and folder git trees
  (code files, sibling docs, PRs, issues, branches), by relation and provenance;
  \texttt{repo\_trees} holds the git tree that references resolve against.
\end{itemize}

\begin{figure}[t]
\centering
\begin{tikzpicture}[
  every node/.style={font=\scriptsize},
  entity/.style={draw, rounded corners=2pt, align=left, inner sep=3pt,
                 text width=3.05cm, fill=blue!4},
  spine/.style={draw, rounded corners=2pt, align=left, inner sep=4pt,
                text width=3.25cm, fill=orange!12, very thick},
  census/.style={draw, rounded corners=2pt, align=left, inner sep=4pt,
                 text width=3.05cm, fill=green!6, thick},
  rel/.style={-{Latex[length=1.6mm]}, thin, gray!70},
  klab/.style={font=\tiny\ttfamily, fill=white, inner sep=1pt, midway}
]
\node[spine] (sf) {\textbf{spec\_files} \tiny(broad census)\\[1pt]\ttfamily file\_url\_sha16 \tiny(PK)\\\ttfamily repo\_name\\\ttfamily spec\_tool, spec\_role\\\ttfamily file\_content};

\node[entity, above left=0.5cm and 1.5cm of sf] (sfc) {\textbf{spec\_file\_commits}\\[1pt]\ttfamily file\_url\_sha16\\\ttfamily commit\_sha, authored\_at};
\node[entity, below left=0.5cm and 1.5cm of sf] (scf) {\textbf{spec\_content\_features}\\[1pt]\ttfamily file\_url\_sha16 \tiny(PK)\\\ttfamily n\_headings, has\_ears \ldots};

\node[census, above right=0.4cm and 1.6cm of sf] (kf) {\textbf{kiro\_files} \tiny(Kiro census)\\[1pt]\ttfamily repo\_name, file\_path\\\ttfamily kind, feature};
\node[entity, below right=0.4cm and 1.6cm of sf] (oaf) {\textbf{openspec\_artifact\_files}\\[1pt]\ttfamily repo\_name, file\_path\\\ttfamily kind, n\_tasks};

\node[entity, left=1.5cm of sf] (sl) {\textbf{spec\_links} \tiny(trace index)\\[1pt]\ttfamily spec\_url\_sha16\\\ttfamily rel, target, provenance};

\node[entity, below=1.6cm of sf] (prs) {\textbf{pull\_requests}\\[1pt]\ttfamily repo\_name, pr\_number\\\ttfamily touches\_code};
\node[entity, right=1.7cm of prs] (prf) {\textbf{pr\_files}\\[1pt]\ttfamily repo\_name, pr\_number\\\ttfamily is\_spec, is\_code};
\node[entity, right=1.7cm of oaf] (ocr) {\textbf{openspec\_code\_refs}\\[1pt]\ttfamily repo\_name, artifact\_path\\\ttfamily code\_path, task\_done,\\\ttfamily exists\_at\_anchor};

\draw[rel] (sf) -- (sfc) node[klab,above]{file\_url\_sha16};
\draw[rel] (sf) -- (scf) node[klab,below]{file\_url\_sha16};
\draw[rel] (sf) -- (sl) node[klab,above]{file\_url\_sha16};
\draw[rel] (sf) -- (oaf) node[klab,below]{repo\_name};
\draw[rel] (sf) -- (prs) node[klab,right]{repo\_name};
\draw[rel] (prs) -- (prf) node[klab,above]{pr\_number};
\draw[rel] (oaf) -- (ocr) node[klab,above]{artifact\_path};
\end{tikzpicture}
\caption{Entity-relationship diagram of the released \dsname{} schema. The
\texttt{spec\_files} spine (orange) is the broad census; its file-keyed
satellites join on \texttt{file\_url\_sha16}. The Kiro census (green) is captured
independently and keyed on \texttt{repo\_name}. The per-tool PR tables (11 tools,
shown once) record the pull requests that change a spec and their full
changesets, flagging each file as spec or code. \texttt{spec\_links} and
\texttt{openspec\_code\_refs} form the traceability index: each spec's links to code,
PRs, issues, and sibling docs, the latter resolved against the repository tree. Plumbing tables
(discovery queues, partition plans, raw staging) are omitted.}
\label{sm:fig:er}
\end{figure}

\begin{table}[t]
\centering
\caption{Condensed data dictionary of the core released tables. The full
dictionary (all tables and columns, with types and provenance notes) ships as
\texttt{DATA\_DICTIONARY.md}.}
\label{sm:tab:dict}
\small
\setlength{\tabcolsep}{4pt}
\begin{tabular}{@{}l r p{12.2cm}@{}}
\toprule
Table & Rows & Purpose / join key \\
\midrule
\texttt{spec\_files} & 470,795 &
  broad census spine: full repo object, file/commit provenance, tool
  attribution, raw content. \emph{PK} \texttt{file\_url\_sha16}. \\
\texttt{spec\_file\_commits} & 780,335 &
  Complete per-file commit history (author, committer, timestamps, message).
  \emph{FK} \texttt{file\_url\_sha16}. \\
\texttt{spec\_content\_features} & 468,307 &
  39 structural and requirement-template features per spec. \emph{PK}
  \texttt{file\_url\_sha16}. \\
\texttt{kiro\_files} & 98,574 &
  Kiro census under \texttt{.kiro/specs}; \texttt{kind} = requirements / design /
  tasks. Key \texttt{repo\_name, file\_path}. \\
\texttt{pull\_requests} & 5,992 &
  Pull requests that modify a spec: state, merge, branch, line counts,
    \texttt{touches\_code}, \texttt{tool}. Key \texttt{repo\_name, pr\_number}.
    Concatenates the eleven per-tool crawl tables; 8 PRs appear in more than
    one, because their repository ships more than one tool, adding 13 rows, so
    the distinct count is 5,979. \\
\texttt{pr\_files} & 348,141 &
  Per-file changeset of each spec-touching PR, each file flagged
  \texttt{is\_spec} / \texttt{is\_code}. \emph{FK} \texttt{repo\_name, pr\_number}.
  Distinct on (\texttt{repo\_name}, \texttt{pr\_number}, path): 346,991. \\
\texttt{openspec\_artifact\_files} & 266,230 &
  OpenSpec change proposals, designs, and task lists, with parsed task counts.
  Key \texttt{repo\_name, file\_path}. \\
\texttt{openspec\_code\_refs} & 435,401 &
  Task-to-code references parsed from OpenSpec \texttt{tasks.md} and resolved
  against the tree at the anchoring commit (\texttt{task\_done},
  \texttt{exists\_at\_anchor}). \emph{FK} \texttt{repo\_name, artifact\_path}. \\
\texttt{spec\_links} & 2,421,323 &
  Census-wide index of typed references (from spec text, commit messages,
  \texttt{tasks.md}, and folder trees): code, sibling docs, PRs, issues, branches,
  by relation and provenance. \emph{FK} \texttt{spec\_url\_sha16}. \\
\bottomrule
\end{tabular}
\end{table}

\phantomsection\label{sm:app:qc}
\subsection*{Appendix B: Construction and Quality Control}
Because GitHub code search caps any query at 1{,}000 results and reports an
unreliable \texttt{total\_count}, both censuses are built by \emph{adaptive size
partitioning}: file-size ranges are recursively split until each leaf is safely
below the cap, and each leaf is paged until a genuinely empty page (short pages,
which GitHub returns silently under rate-limiting, are retried rather than
treated as the end). We checked that the procedure is idempotent (re-running adds
$\approx$0 files) and deterministic (independent fetches converge on the same
set). The reachable set may still be a strict subset of all matching files.

The sweep stored \num{822{,}901} rows across its size partitions, which reduce to
\num{575{,}633} distinct files once deduplicated on the blob URL; of these, \num{470{,}795}
survive a documented \emph{not-a-spec} filter that removes files inside package
caches (\num{97{,}757}) or vendored dependency trees (\num{376}) and specs first
committed before 2024 (\num{6{,}705}). Each filter is a documented, reversible rule
keyed on a flag and reason; the released tables contain the kept set, and any
excluded file can be recovered by re-running the pipeline or re-fetching it from
GitHub.

\paragraph{Maturity and filtering}
A filename census of a practice this young mixes serious projects
with tutorials, template clones, and one-off experiments: only 923 of the
\num{73{,}030} repositories have 100 stars or more. We do not pre-filter this
for participants, because where the line falls depends on the study, but we ship
what is needed to draw it: fork, template, and archived flags on every
repository object, a content hash on every file so that copies of the same
starter template can be collapsed, and per-spec \texttt{is\_tiny},
\texttt{has\_lorem}, and unfilled-placeholder flags that identify scaffolding
nobody ever filled in. Reporting which filter was applied should be part of any
study built on \dsname{}. The two censuses are keyed independently and may
overlap: a team can use Kiro in a repository that also has a
\texttt{spec.md}.

\paragraph{Scope of the PR layer}
The PR layer is a subsample, not a census. The eleven tools are OpenSpec, Spec
Kit, Claude Code, Conductor, Trae, Codex, Agent OS, OpenCode, MoAI, SpecWeave
and Cursor: the tools whose attribution labels existed when the PR crawl ran in
July~2026, not a ranking by adoption. SpecWeave yielded no spec-touching pull
requests, so the released layer holds ten tools, not eleven. Three named tools
labelled after that date (\texttt{.specs/}, ZenFlow and Kitty Specs, together
12 eligible repositories) and the Kiro census (173 eligible repositories) have
no PR layer at all. Within each of the eleven we ran discovery on every repository with at
least ten stars, the selection rule for the PR layer: 963 repository--tool targets
over 949 distinct repositories (a repository can adopt more than one tool),
dominated by OpenSpec (524) and Spec Kit (305). Of these, 581 contain at least one
spec-touching pull request and are included in the released layer; the rest
committed their specs directly to a branch or did not have any qualifying PR.
Results from this layer
describe those repositories in detail and should not be extrapolated to the full
\num{73{,}030}-repository census without care. Within a sampled repository
the sweep is exhaustive for the PRs it targets: every PR that touches a spec
file is captured with its changeset, complete up to GitHub's 3{,}000-file-per-PR
API cap, at which a small number of very large PRs are truncated. Discovery is
commit-based (a PR is included when a spec file's commit is associated with it
through GitHub's \texttt{/commits/\{sha\}/pulls}), while the changeset is the
PR's net diff, so the two occasionally disagree: five of the \num{5{,}979} PRs
do not have any spec file in their diff, the truncated ones plus two whose spec
commit is in shared or ancestor history. By construction, a PR that does not
touch any spec file never reaches the sweep, even when it implements one;
recovering those means enumerating a repository's full PR history from the API
and filtering it, which the released repository list and PR numbers make
possible but which we have not done ourselves.

\paragraph{The co-change heuristic and its limits}
How a spec actually becomes code is not directly observable, and we do not
answer that question here. The PR layer instead applies a deliberate heuristic that
participants can inspect: \emph{a pull request that touches a spec file and
also changes source files is treated as implementing that spec}. Co-change is
common in the PR layer (81.2\% of spec-touching PRs also modify code, in the same
reviewed changeset), so it is a reasonable place to start, but it remains an
assumption. A PR may touch a spec for unrelated reasons or
implement it only partially. A workflow that merges a spec on its own and
implements it in later PRs is invisible to co-change altogether. The complete
changeset of every spec-touching PR, with branch names, authors, and creation
and merge timestamps, lets participants tighten or replace the heuristic (by
file-path proximity, commit and merge timing, task-list checkbox transitions, or
branch and title references to a named spec). The traceability index records
cases co-change misses: the PRs and issues a spec names appear whether or
not those PRs touch a spec file. Reconciling the two channels, and measuring how
much each one covers, is itself one of the research questions this dataset
opens (Sec.~\ref{sm:sec:rq}, item~3).

\phantomsection\label{sm:app:vocab}
\subsection*{Appendix C: Controlled Vocabularies}
The values each categorical column takes, with row counts in \dsname{}~v1.0.

\noindent\textbf{\texttt{spec\_files.spec\_role}} (lifecycle role):
archived (\num{156{,}936}), na (\num{137{,}779}), living (\num{73{,}253}),
feature (\num{49{,}370}), change\_proposal (\num{41{,}806}), config
(\num{11{,}651}). \emph{living}/\emph{change\_proposal}/\emph{archived} are the
OpenSpec lifecycle; \emph{feature} is a Spec Kit numbered feature; \emph{config}
is a tool scaffold file; \emph{na} is everything else.

\noindent\textbf{\texttt{spec\_links.rel}} (what a reference points to):
code (\num{1{,}278{,}828}), sibling (\num{863{,}445}), pr (\num{151{,}563}),
ref (\num{62{,}267}), branch (\num{42{,}970}), issue (\num{22{,}139}),
anchor (111).

\noindent\textbf{\texttt{spec\_links.provenance}} (how the reference was found):
tasks (\num{1{,}194{,}660}, from an OpenSpec \texttt{tasks.md}), tree
(\num{937{,}815}, from listing the repository git tree), commit\_msg
(\num{231{,}284}, from the spec's commit messages), branch\_header
(\num{42{,}970}, the \texttt{Feature Branch:} line), content (\num{14{,}594},
parsed from the spec body).

\noindent\textbf{\texttt{kiro\_files.kind}}:
requirements (\num{32{,}972}), design (\num{32{,}123}), tasks (\num{31{,}592}),
bugfix (\num{1{,}539}), other (348).

\noindent\textbf{\texttt{openspec\_artifact\_files.kind}}:
proposal.md (\num{93{,}307}), tasks.md (\num{93{,}061}), design.md
(\num{79{,}459}), plan.md (245), research.md (78), data-model.md (41),
quickstart.md (39).

\noindent\textbf{PR-file flags} (\texttt{pr\_files}, per
changed file): \texttt{is\_spec}, \texttt{is\_code}, \texttt{is\_specdir}
(booleans); \texttt{change\_status} (added / modified / removed). The PR-level
\texttt{touches\_code} is 1 when a PR changes at least one code file outside the
spec directory.

\phantomsection\label{sm:app:features}
\subsection*{Appendix D: Content and Structural Features}
Each spec's Markdown is parsed into 39 features (\texttt{spec\_content\_features},
one row per spec, keyed by \texttt{file\_url\_sha16}), grouped below. Count
columns are integers; marker columns are 0/1.

\begin{itemize}
\RaggedRight
\item \textbf{Size:} \texttt{n\_bytes}, \texttt{n\_lines}, \texttt{n\_words}.
\item \textbf{Structure:} \texttt{n\_headings}, \texttt{max\_heading\_depth},
  \texttt{n\_code\_fences}, \texttt{n\_mermaid}, \texttt{n\_tables},
  \texttt{n\_links}, \texttt{n\_images}, \texttt{n\_list\_items},
  \texttt{n\_checkboxes}, \texttt{n\_checked}.
\item \textbf{Requirement syntax (counts):} \texttt{n\_shall}, \texttt{n\_gherkin},
  \texttt{n\_scenario\_blocks}, \texttt{n\_delta\_headers}.
\item \textbf{Section / requirement markers (0/1):} \texttt{has\_ears},
  \texttt{has\_gherkin}, \texttt{has\_user\_story}, \texttt{has\_requirements},
  \texttt{has\_acceptance\_criteria}, \texttt{has\_user\_scenarios},
  \texttt{has\_non\_goals}, \texttt{has\_out\_of\_scope}, \texttt{has\_overview},
  \texttt{has\_success\_criteria}, \texttt{has\_edge\_cases},
  \texttt{has\_open\_questions}, \texttt{has\_dependencies}, \texttt{has\_purpose},
  \texttt{has\_mandatory\_marker}.
\item \textbf{Quality / completeness:} \texttt{n\_needs\_clarification},
  \texttt{n\_todo}, \texttt{is\_tiny}, \texttt{has\_unfilled\_placeholder},
  \texttt{has\_lorem}.
\item \textbf{Language:} \texttt{lang} (detected natural language),
  \texttt{content\_family} (template family, e.g.\ openspec / speckit).
\end{itemize}

\phantomsection\label{sm:app:tools}
\subsection*{Appendix E: Full Tool Attribution}
Table~\ref{sm:tab:fulltools} lists every attribution category in the broad census:
all 17 named tools, the 4 interpretable path buckets (groupings of unattributed
files, not tools), and the separate Kiro census, with file and repository counts.
caffeine.ai has the most repositories but is an app \emph{generator} whose
repositories are largely auto-generated single apps; among developer-adopted
tools the leaders are Kiro, Spec Kit, and OpenSpec.

\begin{table}[H]
\centering
\caption{Every attribution category in \dsname{}~v1.0 (kept spec set).}
\label{sm:tab:fulltools}
\small
\begin{tabular}{@{}lrr@{}}
\toprule
Tool / bucket & Files & Repos \\
\midrule
OpenSpec       & \num{274{,}955} & \num{8{,}926} \\
Spec Kit       & \num{54{,}640}  & \num{10{,}619} \\
caffeine.ai    & \num{14{,}262}  & \num{13{,}749} \\
Conductor      & \num{7{,}752}   & \num{936} \\
Trae           & \num{2{,}004}   & \num{434} \\
opencode       & \num{1{,}963}   & \num{66} \\
dot\_specs     & \num{1{,}619}   & \num{250} \\
Claude Code    & \num{1{,}351}   & \num{833} \\
Agent OS       & \num{1{,}278}   & \num{237} \\
MoAI-ADK       & \num{1{,}156}   & \num{63} \\
specweave      & \num{963}       & 4 \\
kitty\_specs   & \num{840}       & \num{67} \\
codex          & \num{550}       & \num{69} \\
zenflow        & \num{234}       & \num{64} \\
gsd            & \num{196}       & \num{181} \\
cursor         & \num{80}        & \num{65} \\
kilocode       & 3               & 3 \\
\midrule
\multicolumn{3}{@{}l}{\emph{Path buckets (not tools):}} \\
adhoc\_other   & \num{38{,}575}  & \num{10{,}239} \\
specs\_generic & \num{31{,}725}  & \num{3{,}301} \\
docs\_embedded & \num{19{,}437}  & \num{8{,}655} \\
root\_bare     & \num{17{,}212}  & \num{17{,}104} \\
\midrule
\textbf{Kiro} (separate census) & \num{98{,}574} & \num{12{,}910} \\
\bottomrule
\end{tabular}
\end{table}

\endgroup

\section{Repository Mining Details}\label{app:mining}

This section expands on the collection, attribution, sampling, enrichment,
and judging procedures summarized in Section~\ref{sec:method-mining}.

\begin{figure}[h]
  \centering
  \includegraphics[width=0.8\textwidth]{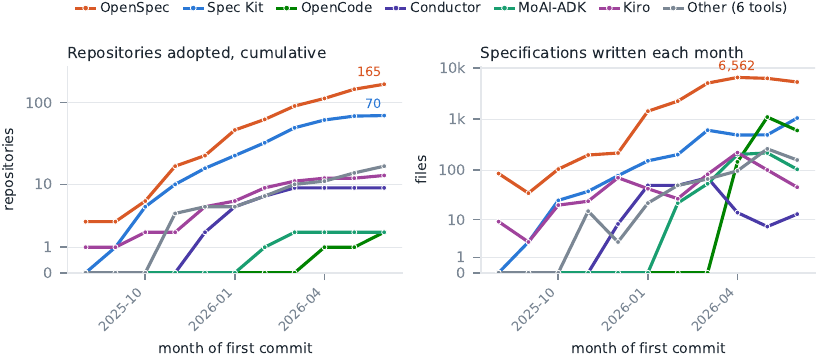}
  \caption{Adoption by tool within the repository sample: cumulative
    repositories (left) and specs written per month (right), on log axes. A
    repository adopts a tool on the first-commit date of its earliest
    file attributed to that tool; ten repositories use more than one tool, so
    the left panel sums to 279 adoptions rather than 266 repositories and its
    curves are not a partition. Every spec carries one attribution, so the
    right panel is. Shown are the five tools with the most files and Conductor, which ranks fourth by repositories adopted. The remaining six are pooled. July 2026 is omitted, since collection ended on 2026-07-03 and the
    partial month would read as a decline; it holds 2,122 specs and one
    Spec~Kit adoption, which is why the left panel ends at 70 rather
  than 71.}
  \label{fig:timeline}
\end{figure}

\textit{Code search.} GitHub code search caps a query at 1,000 results.
Because a file has exactly one size, disjoint size ranges partition the
matches, so we query each range separately
(\texttt{filename:spec.md size:1000..1999}), recursively halving any range
that comes back full, and page through each until a page comes back empty.
The union of these ranges is every file the index exposes for that filename,
which need not be every such file on the platform. The collection produced
822,901 records, corresponding to 575,633 spec files after removing
duplicates of the same blob URL. We excluded files held in package caches
(97,757), in vendored dependency trees (376), or first committed before 2024
(6,705). The published corpus ships the included specs only.

\textit{Tool attribution.} We match each file's path against the
directory patterns of 18 named tools in a fixed order, most specific first,
and take the first match: OpenSpec writes under \texttt{openspec/}, Spec~Kit
under \texttt{specs/NNN-feature/} or \texttt{.specify/}, Kiro under
\texttt{.kiro/specs/}, and so on. OpenSpec nests \texttt{specs/} inside
every change, so without that order its files would fall to the generic
\texttt{specs/} bucket. A file matching no tool falls into one of four
buckets: a \texttt{specs/\textit{name}/spec.md} layout carrying no tool
marker, a spec inside a \texttt{docs/} tree, a bare \texttt{spec.md} at the
repository root, and everything else. These buckets record that no tool
marker was found, not that no tool was used: a project may follow a
spec-driven workflow of its own design, or use a tool with a non-standard
layout.

\textit{Lifecycle roles.} OpenSpec keeps one capability spec for each capability under \texttt{openspec/specs/}. A change is not edited into that file but written as change specs under
\texttt{openspec/\allowbreak changes/\allowbreak\textit{<change>}/\allowbreak
specs/\allowbreak\textit{<capability>}/},
which names the capability it modifies, and moves to
\texttt{openspec/\allowbreak changes/\allowbreak archive/} when complete. A
further 135 files under
\texttt{openspec/} match none of the three roles and form a fourth class.
Spec~Kit creates every spec for a single change but does not record whether
the spec was later updated as a living spec, and the remaining tools encode
no distinction, so we report roles for OpenSpec only and do not treat role
vocabularies across tools as comparable.

\textit{Sample construction.} The five filters are conjunctive, so their
order changes only the intermediate yields in Table~\ref{tab:corpus};
attribution comes first because it is the one condition on the artifact
rather than on the repository that holds it. We count commits by distinct
SHA rather than summing per-file commit counts, because one commit often
modifies several specification files. Kiro passes through the same cascade:
12,497 of its 12,910 repositories hold at least one
\texttt{requirements.md}, 285 of those clear ten stars, 171 survive
licensing, and 13 reach ten distinct commits, contributing 654 requirements
documents. None of those 13 repositories also holds a \texttt{spec.md}, so
they add to the 253 repositories and 36,236 spec files that survive the
cascade from the filename search, giving 266 repositories and 36,890 files.

Twelve of the eighteen named tools survive the filtering (OpenSpec,
  Spec~Kit, OpenCode, Kiro, MoAI-ADK, Codex, spec-kitty, Conductor,
\texttt{.specs}, Trae, Claude~Code, and Cursor). The other six are absent
because no repository using them clears the star, license, and commit
thresholds. OpenSpec accounts for 78.4\% of the sample's specs, writing
6,562 in its busiest month against 7,953 from all other tools across the
entire period.

\textit{Enrichment.} For each spec, we record the commit history
retrieved for its path, its lifecycle role where the tool encodes one, and
39 parsed structural features, including document size, heading structure,
requirement syntax such as EARS~\cite{mavin2009ears} and Gherkin, section
markers, and completeness signals such as unfilled placeholders. The full
feature list is included with the dataset.

\textit{Pull requests.} We include a PR in the corpus when GitHub
associates a spec commit with it through the \texttt{/commits/{sha}/pulls}
endpoint, which also catches rebase merges that leave no marker in the
commit message. For each included PR, we record the net diff and classify
every changed file as a spec or code file. We apply this procedure to the
eleven tools for which we built a per-tool crawl, taking every repository
using one of them that has at least ten stars and a license: 949 distinct
repositories, eleven of which use more than one tool. Querying the 15,726
spec commits those repositories hold yields 5,979 distinct spec-modifying
PRs from 581 of them. Two tools present in the repository sample,
\texttt{.specs} and spec-kitty, have no such crawl and are absent from this
layer. Where a PR modifies both a spec and source code,
we read the co-occurrence as evidence that the PR implements that spec. A PR
may touch a spec for an unrelated reason, and a workflow that merges a spec
first and implements it later is not captured by co-change at all.

\textit{Heading parsing.} For each file in the sample we fetch the raw
Markdown pinned to the file's last commit, so the text matches the metadata
snapshot already recorded; all 36{,}890 files resolved. We parse each
document with a CommonMark-compliant parser rather than matching lines that
begin with \texttt{\#}, because 7.3\% of files contain fenced code in which
such lines are not headings. Parsing yields 527{,}981 headings across the
36{,}881 files that have at least one. Appendix~\ref{app:headings} gives the
parser corrections and normalization rules.

\textit{LLM judging.} Before sampling the 1{,}000 documents for the
content and work classifications, files with byte-for-byte identical text
are collapsed to a single representative. The sample is drawn equally across
the 13 document types so that each type is estimated with comparable
precision, and estimates weight each type by its number of files in the repository sample. To prevent the model from inferring
labels from recognizable templates or metadata, we provide only the document
text, excluding repository, path, tool, and role information, and require a
verbatim supporting quote for every assigned label. After Markdown
normalization, the cited quotes match the source text for 98.3\% of content
labels and 99.4\% of work labels.

\textit{Author placement.} A commit whose author carries no linked
account is matched to one when the same name or address appears with a login
on another commit in the same repository, and is otherwise left unplaced.

\section{SDD Tool Documentation}\label{app:tools}

Table~\ref{tab:tools} compares what the five dedicated SDD tools in the
repository sample prescribe, and what the sample shows for each tool's files.
OpenCode is omitted because it is a coding agent and nearly all its files come
from one repository. We read each open-source tool's documentation as of its
last release or commit before collection ended on 3 July 2026 (OpenSpec
v1.5.0, Spec~Kit and MoAI-ADK main branch, Conductor v0.4.1). Kiro is closed
source and its documentation is not versioned, so its column reflects the
documentation as accessed in October 2026. Most of MoAI-ADK's files come from
the vendor's own repository, and the Spec~Kit column reports its
\texttt{specs/} files, not its \texttt{.specify/} scaffold. The tools also differ in how many repositories try them and how many keep using them. Kiro appears in 12,497 repositories of the corpus and 13 of the sample, Spec~Kit in 10,619 and 71, and OpenSpec in 8,926 and 165.

\begin{table}[!htb]
  \caption{What five SDD tools prescribe, and what the repository sample shows
  for each tool's files.}
  \label{tab:tools}
  \scriptsize
  \setlength{\tabcolsep}{3pt}
  \begin{tabular}{@{}>{\raggedright\arraybackslash}p{0.13\textwidth}
    *{5}{>{\raggedright\arraybackslash}p{0.158\textwidth}}@{}}
    \toprule
    & OpenSpec & Spec~Kit & Kiro & Conductor & MoAI-ADK \\
    \midrule
    Spec scope & Capability spec; change specs per change & Change request (numbered feature) & Feature & Change request (track) & Change request \\
    \addlinespace[3pt]
    Agent support & About 30 agents & About 34 agents & Kiro's own IDE and CLI & Gemini CLI & Claude Code \\
    \addlinespace[3pt]
    Implementation detail & Design file; excluded from spec & Plan file; excluded from spec & Design file & Plan file & Plan file \\
    \addlinespace[3pt]
    Approval & None required & User runs each step & Between phases & Spec, plan, each phase & Spec, before implementation \\
    \addlinespace[3pt]
    Maintenance
    & Living spec and accumulated change record
    & Team's choice 
    & Living spec
    & Tracks archived or deleted after review
    & Spec status updated after implementation \\
    \addlinespace[3pt]
    Verification & On request, per change
    & On request, per change request
    & Optional tests derived from requirements
    & Tests written first; manual check per phase
    & Tests written first; quality gates \\
    \midrule
    Median words per spec (observed) & 527 capability specs, 334 open change specs, 292 archived changes & 1,286 & 877 & 290 & 1,745 \\
    \addlinespace[3pt]
    Implementation detail in a section or more (observed) & 75\% capability specs, 55\% open change specs, 40\% archived changes
    & 88\%
    & 75\%
    & 81\%
    & 95\% \\
    \addlinespace[3pt]
    Single recorded commit (observed) & 47\% capability specs, 84\% open change specs, 95\% archived changes
    & 74\%
    & 74\%
    & 92\%
    & 22\% \\
    \bottomrule
  \end{tabular}
\end{table}

\section{Repository Sample Results in Detail}\label{app:specmine-results}

This section gives the results from the repository sample that the findings
summarize.

\textit{Implementation detail (Finding~1).} Nearly every spec we judged
states requirements, and 77.6\% also state how the software is to be built; in
58.2\% that material occupies a section or more. Of the specs with
implementation detail, 75\% name classes, functions, file paths, data
structures, or library versions, and 71\% describe components, interfaces, or
data models. Specs under Spec~Kit's \texttt{.specify/}, MoAI-ADK, and
\texttt{.specs} devote a section or more to implementation in at least 94\% of
files, against 40\% of OpenSpec's archived changes and 55\% of its open
change specs. OpenSpec ranks lowest and is also the tool that writes design into
a sibling document, so part of the gap reflects where a tool puts
implementation detail rather than whether its users write any. We judged each
document on its own text and did not check any sibling file.

\textit{Section headings (Finding~1).} A section either belongs to a
type's template or is nearly absent from that type's files. Pairing the 13 document types with the 40 sections that head a majority of files in at least one type gives 520 type--section pairs. Figure~\ref{fig:headings} shows the pairs for the 10 sections that recur across types. Of all 520 pairs, 52\% are exactly zero and
82\% fall below a tenth, while those at half or more have a median of 66\%.
Of OpenSpec capability specs, 81\% head \emph{requirements}; of its change
specs, 83\% head \emph{added requirements}, the delta vocabulary the tool
writes into its change template; and 87\% of Conductor files head
\emph{acceptance criteria}. Half of Spec~Kit's \texttt{specs/} files keep the
literal \texttt{(mandatory)} annotation it writes into its scaffold. Across the
13 document types, 40 sections reach a majority and 30 appear in exactly one.
Nine of the other ten cross a tool boundary, and all nine are general
requirements vocabulary the tools inherited. Even these concentrate in a few
types: \emph{functional requirements} heads 64--97\% of files in the four types
that use it and at most 9\% in the other nine. Document types from the same
tool also differ. OpenSpec's capability specs share no majority section with
its change records, and Spec~Kit's two locations share one. Shared headings do
not imply identical content or an unchanged workflow. Implementation detail
can fit within existing sections or accompanying design and plan files, and
agents, which drafted most specs (Appendix~\ref{app:generic}), may preserve
template instructions while following customized guidance.

\textit{Purpose of specs (Finding~2).} Following Watanabe et
al.~\cite{watanabe2025agentic}, we labeled the purpose of 1{,}000 specs and weighted each
document type by its number of files; a spec may carry several labels. An estimated 82.0\% carry the \emph{feat} label, against
16.5\% for \emph{fix}, 24.3\% for \emph{refactor}, and 11.6\% for \emph{perf}.
The share of \emph{feat} runs from 51\% to 97\% across the 13 document types.
The labels describe purpose and do not establish whether task size or urgency
explains the pattern.

\textit{Authorship and review (Finding~3).} Two Kiro files have no recorded history. Of the other 36{,}888 spec files, 98.3\% have a single author, and in 63\% of the 266
repositories only one person has ever touched a spec. Of 1{,}731 inline
comments anchored to a spec, 68.5\% were written by bots, and 210 of the 266
repositories have no such comment. Of the 545 comments written by people, 79\%
reply within a thread someone else opened, leaving 112 comments that start a
thread, spread across 31 repositories; the second reader of a spec, where there
is one, is usually a bot. No spec commit linked to a pull request came from a
contributor identified as contributing to that repository for the first time.

\textit{Revision (Finding~4).} Within OpenSpec, 95.4\% of archived changes
have one commit, against 84.2\% of open change specs and 47.4\% of capability
specs, 30.2\% of which have three or more. Counting files rather
than commits, archived changes outnumber capability specs by 19{,}067 to
6{,}644. Which unit grows follows from the lifecycle stage and does not by
itself show that repositories favor accumulating records over maintaining
living specs. MoAI-ADK is the exception among tools: 22.3\% of its specs have
one commit and 60.3\% have three or more.

\textit{How projects divide their specs (Finding~5).} A project that keeps
one spec per feature should hold a few long documents; a
project that writes a spec for each change should hold many short ones, since
every change adds another. We exclude OpenSpec's change records, whose number
reflects how long a project has used the tool rather than how it scopes a
spec. The proxy cannot separate a system with many features from one with many changes, so we read it as consistent with the interview accounts rather
than as independent confirmation.

\textit{Spec authorship (Finding~6).} Placing each author against their
repository's contributor list, repository owners and the three most frequent
committers account for 90\% of the 13{,}640 spec commits while making up 60\%
of the people who write them. Three people, with seven commits between them,
wrote a spec for a repository they had not otherwise contributed to. A
repository records only the specs a project chose to publish
(Section~\ref{sec:threats}).

\section{Heading Extraction}\label{app:headings}

After parsing, we apply three corrections specific to this corpus. A leading
byte-order mark is stripped, because it hides the front matter and turns the
YAML body into a heading. Headings that are empty once inline HTML comments are
removed are dropped. Setext headings whose text is a YAML key are suppressed;
such headings occur where a template embeds a delimited block mid-document.

Each heading is kept as written and in a normalized form, with links, images,
inline code, HTML, emphasis, anchors and leading numbering removed and the text
lowercased, so that ``10. Out of Scope'' and ``2. Out of scope'' aggregate with
``Out of Scope''. Without normalization, the sections a template supplies would split
across spellings:
\emph{out of scope} is written 36 different ways across the corpus, and
\emph{success criteria} 29.

A line-matching count over the same files differs from the parser on 1.1\% of
them, and 91\% of those contain a fenced code block. Of the 36{,}890 files,
nine contain no heading at all; Kiro contributes 12{,}916 of the 527{,}981
headings.

\section{Judge Prompts}\label{app:prompts}

Each prompt is sent with the document text appended under a \texttt{DOCUMENT:}
delimiter. A document longer than 6{,}000 words keeps its first 3{,}000 and last
1{,}000 words, with the elision marked. Line breaks below are set for this page.

\subsection{Spec content}\label{app:prompt-content}

\begin{footnotesize}\begin{verbatim}
You are reading a software specification document from a public GitHub
repository. A developer wrote it, or an AI coding agent wrote it on
their behalf, to describe work to be done.

Decide two things independently: does the document state REQUIREMENTS,
and does it state IMPLEMENTATION detail. Both can be present. Either can
be absent.

REQUIREMENTS - what the software must do, and when the result is correct
  problem      the problem being solved, the objective, or the rationale
  behaviour    what the system must do: user stories, use cases, user
               journeys, functional statements, flows and conditions
  acceptance   conditions under which the result is correct: acceptance
               criteria, edge cases, failure scenarios, expected outputs
  constraints  non-functional requirements, performance or security
               limits, guardrails, or items declared out of scope
  stakeholders who the work is for, business context, or affected users

IMPLEMENTATION - how the software is to be built
  architecture components, modules, interfaces, API surfaces, data
               models, or design decisions and their trade-offs
  plan         task breakdown, sequencing, phases, migration steps, or
               the choice of a technology, library, framework or
               datastore
  code         named classes or functions, file paths to create or edit,
               data structures, algorithms, configuration values,
               library versions, or code to write
  conventions  coding standards the implementer must follow, or internal
               API and domain detail they would not otherwise have

RULES
1. The two dimensions are independent. Decide each on its own. A
document that
   is entirely implementation detail has implementation=1 and
      requirements=0; a
   document that states behaviour and nothing else has the reverse.
2. Acceptance criteria are REQUIREMENTS, not implementation. They state
when a
   result is correct, not how to produce it. Count testing as
          implementation
   only where the document prescribes the mechanics, such as the test
        framework
   to use or the file to put tests in.
3. Judge on POSITIVE EVIDENCE in the text. A heading with nothing under
it is
   not evidence. Neither is a passing mention: "unlike the old
       scheduler" is not
   architecture unless the document specifies something about the
                architecture.
4. Judge the document in front of you. Do not credit it for content a
sibling
   file is likely to hold, and do not penalise it for deferring detail
   elsewhere. A link or reference to another document is not that
              content.
5. Code blocks are not automatically implementation. A block showing an
expected
   API response is acceptance; a block showing the code to write is
       code.
6. Some documents are unfilled templates, or describe a system as it
already
   stands rather than work to be done. Set is_spec=false for those, and
          still
   label both dimensions on whatever content is present.
7. Documents may be in any language. Judge them the same way and answer
in
   English.
8. For every category you mark present, quote at most 20 words copied
EXACTLY
   from the document. If you cannot quote evidence, the category is
        absent.
9. Also judge HOW MUCH of the document each dimension occupies:
     none         absent
     incidental   a line or two in passing
     substantial  a section or several statements, but not the bulk
     dominant     most of the document

Return one JSON object and nothing else:

{"is_spec": true or false,
 "requirements": 0 or 1,
 "implementation": 0 or 1,
 "requirements_extent": "none" or "incidental" or "substantial" or
  "dominant",
 "implementation_extent": "none" or "incidental" or "substantial" or
  "dominant",
 "requirements_categories": ["behaviour", ...] or [],
 "implementation_categories": ["architecture", ...] or [],
 "evidence": {"behaviour": "exact quote", "architecture": "exact quote",
  ...},
 "dominant": "requirements" or "implementation" or "balanced" or
  "neither",
 "confidence": "high" or "medium" or "low",
 "note": "one short sentence; flag anything ambiguous"}
\end{verbatim}\end{footnotesize}

\subsection{Kind of work}\label{app:prompt-work}

\begin{footnotesize}\begin{verbatim}
You are given a specification document written for some software
engineering task. It comes from a public GitHub repository and was
written by a developer, or by an AI coding agent on their behalf, to
describe work to be done. Your job is to infer what kind of work it
specifies.

Use the taxonomy below. It has two INDEPENDENT dimensions. A document
may carry MULTIPLE labels on each dimension, because one specification
often calls for several kinds of work at once (for example: add a
feature, add tests for it, and update the README).

PURPOSE TYPES - what the work does to the behaviour of the code
  fix       Code changes that fix bugs and faults within the codebase.
  feat      Code changes that introduce new features to the codebase,
            encompassing both internal and user-oriented features.
  refactor  Code restructuring without changing its behavior, aiming to
            improve maintainability.
  style     Non-functional code changes that improve readability or
            consistency.
  perf      Code changes that improve performance, such as enhancing
            execution speed or reducing memory consumption.

OBJECT TYPES - what kind of artifact the work touches
  docs      Updates to documentation or comments, such as README edits,
            typo fixes, or API docs improvements.
  test      Additions or modifications to test files, including new test
            cases or updates to existing tests.
  ci        Changes to CI/CD workflows and configurations, e.g.
            ".travis.yml" and ".github/workflows".
  build     Changes to build configurations (e.g. Maven, Gradle, Cargo).
  chore     Project-wide housekeeping tasks such as dependency bumps,
            version increments, and miscellaneous cleanup.

RULES
1. The two dimensions are independent. Decide each separately. A purpose
label
   must not suppress an object label, or the reverse.
2. Assign a label only on POSITIVE EVIDENCE in the text. Do not infer
"test"
   because good practice implies testing; assign it only if the document
   actually calls for test work. When in doubt, leave it out.
3. If no purpose applies, return ["none"] for purpose; likewise for
object. A
   document may legitimately have purpose labels and no object labels,
            or the
   reverse.
4. Judge what the document SPECIFIES, not what it merely mentions.
"unlike the
   old caching layer, which was slow" is not perf unless the document
       asks for
   a performance improvement.
5. Some documents describe a system as it already is, or are unfilled
   templates, rather than specifying work. Set is_spec_of_work to false
              for
   those and return ["none"] on both dimensions.
6. Documents may be in any language. Classify them the same way
regardless of
   language, and answer in English.
7. For EVERY label you assign, quote a span of at most 20 words copied
EXACTLY
   from the document that justifies it. If you cannot quote evidence for
        a
   label, do not assign that label.

Return one JSON object and nothing else:

{"is_spec_of_work": true or false,
 "purpose": ["feat", ...] or ["none"],
 "object": ["test", ...] or ["none"],
 "evidence": {"feat": "exact quote from the document", ...},
 "primary": "the single label that best captures the document, or none",
 "confidence": "high" or "medium" or "low",
 "note": "one short sentence; flag anything ambiguous"}
\end{verbatim}\end{footnotesize}

\section{Reddit Corpus}\label{app:reddit}

Table~\ref{tab:reddit-dist} lists the communities searched and the number of
records taken from each (Section~\ref{sec:filterproc}). Reddit's search
interface reports ranked results rather than counts, so we report the records
the search returned rather than the size of each community.

\begin{table}[h]
  \caption{Communities searched, with records taken.}
  \label{tab:reddit-dist}
  \footnotesize
  \begin{tabular}{@{}l r@{}}
    \toprule
    Community & Records \\
    \midrule
      \multicolumn{2}{@{}l}{\emph{SDD frameworks}} \\
      \quad r/SpecDrivenDevelopment & 140 \\
      \quad r/BMAD\_Method & 100 \\
      \quad r/speckit & 15 \\
      \quad r/kiroIDE & 31 \\
      \quad r/openspec & 17 \\
      \multicolumn{2}{@{}l}{\emph{AI coding agents}} \\
      \quad r/ClaudeCode & 119 \\
      \quad r/ClaudeAI & 60 \\
      \quad r/codex & 60 \\
      \quad r/cursor & 60 \\
      \quad r/GithubCopilot & 120 \\
      \quad r/google\_antigravity & 60 \\
      \quad r/opencodeCLI & 120 \\
      \multicolumn{2}{@{}l}{\emph{AI-assisted development}} \\
      \quad r/vibecoding & 119 \\
      \quad r/ArtificialInteligence & 59 \\
      \quad r/AI\_Agents & 120 \\
      \quad r/PromptEngineering & 59 \\
      \quad r/AIAssisted & 25 \\
      \quad r/ChatGPTCoding & 60 \\
      \quad r/OnlyAICoding & 24 \\
      \multicolumn{2}{@{}l}{\emph{General SE}} \\
      \quad r/programming & 49 \\
      \quad r/software & 10 \\
      \quad r/ExperiencedDevs & 60 \\
      \quad r/softwaredevelopment & 10 \\
      \midrule
      Total & \num{1{,}497} \\
    \bottomrule
  \end{tabular}
\end{table}

\section{Shared Practices}\label{app:generic}

This section reports practices that most participants shared and that the
findings summarize only briefly. Tool makers are P1, P2, P6, P9, P11, P13, P20, and P21.

\textit{Spec format.} Specs were Markdown documents. A few participants
supplemented Markdown with OpenAPI definitions (P16), UML diagrams (P6), or
HTML for user-interface specs (P19). Some followed a fixed template or schema
of their own (P2, P6, P9, P16, P17, P18, P19), some used the template of an
SDD tool (P4, P5, P8, P10, P11, P12, P13, P15, P20), and some wrote free-form
Markdown (P1, P3, P7, P14, P21). Some considered format secondary (P3, P9,
P11, P18, P21), whereas others considered structure important for agent
reliability (P2, P6, P17, P19). Structured notation such as Gherkin was used
mainly for acceptance criteria and end-to-end tests rather than for the spec
as a whole (P9, P12, P17, P21).

\textit{Modularity and decomposition.} No participant advocated a single,
monolithic spec; P8 had started from a spec base ``completely written from
one single spec file'' and moved away from it. Nearly all participants
decomposed specs vertically, from a product requirement, epic, or roadmap to
specs and from specs to plans and tasks, either through the requirements,
design, and task documents of SDD tools (e.g., P5, P13, P20) or through their
own hierarchies, such as projects, milestones, and tasks (P7) or epics and
user stories (P12). Most tool users who had links between specs at the same
level, or from specs
to code, obtained them from tools (P4, P5, P8, P15, P16), and tool makers
built them into their own tools (P2, P11, P20); P7 and P10 lacked such links
and missed them.

\textit{Where specs come from.} Specs were built from upstream
requirements artifacts produced by others in the organization, such as PRDs,
epics, and user stories (P4, P6, P9, P10, P12, P16); elicited through
conversation or research with the agent, using interview-style skills,
brainstorming, or research prompts (P1, P3, P5, P7, P8, P11, P14, P15, P17,
P18, P19, P21); or reverse-engineered from existing code in brownfield
projects (P2, P6, P8, P10, P21).

\textit{Authoring and refinement.} Agents drafted the specs of nearly all
participants, while participants supplied the intent and reviewed the result.
Humans wrote specs themselves mainly for the most uncertain work (P3, P10,
P14). Participants refined specs iteratively, either until the spec settled
and before implementation (P5, P11, P12, P16, P19) or throughout
implementation, letting the spec evolve or discarding it as implementation
revealed what was needed (P1, P2, P6, P9, P13, P20, P21).

\textit{Spec review.} Participants reviewed specs in full (P2, P6, P11,
P12, P13, P16, P19), read them partially or skimmed them (P4, P5, P7, P8,
P14, P15, P20), or delegated review mainly to agents (P3, P9, P10, P17, P18).
Team review occurred only in organizational settings: by requirements
engineers and stakeholders (P6), a more senior engineer (P10), a sectioned
review by several people (P16), a tech lead who approves the spec (P19), or,
among the users P20 described, whole teams reviewing specs like an RFC.

\textit{Workflow types.} Participants' workflows differed in where a human
checked the agent's work. In a \emph{human-gated, per-change} workflow, the
participant moved through the phases of an SDD tool for each change and
approved each phase, sometimes after only skimming it, and some also reviewed
the resulting code (P4, P5, P10, P12, P13, P15, P16). In a \emph{spec-gated,
then autonomous} workflow, the participant approved the spec and let agents
implement it with little supervision, relying on tests, CI gates, and
reviewer agents (P2, P3, P7, P8, P9, P11, P14, P17, P19, P21). In a
\emph{system-level, design-first} workflow, the participant specified the
whole system before writing code, sometimes over months (P1, P6, P18). Tool
users were divided between the first two types, whereas tool makers mostly
followed the second or third.

\textit{Tests.} Most participants verified code mainly through end-to-end
or acceptance tests (P3, P4, P6, P9, P12, P13, P14, P16, P17, P18, P19, P21)
rather than mainly unit tests (P2, P7, P15). Humans defined the
behavior-level tests that agents implemented (P4, P9, P12, P17, P21),
reviewed the tests that agents wrote (P5, P6, P12), or had a separate team
test the implementation (P16).

\textit{Tools.} Participants used OpenSpec (P5, P7, P8, P9, P10, P13, P15,
P20), Spec Kit (P2, P10, P11, P16), Superpowers (P4, P7, P19), Kiro (P12,
P19), and BMAD (P17), often in combination. Two participants used a tool with
its defaults (P5, P15), tool users mostly extended an existing tool (P4, P7,
P8, P10, P12, P16, P17, P19), and the remaining participants, including all
tool makers, built their own tools or workflows. Participants customized because
existing tools did not fit their approach (P2, P9, P11, P13,
P20), their organization (P4, P6, P10, P12, P19), or their personal workflow
(P7, P17, P21). Tool switches moved toward lighter, more customizable, or
more open tools (P7, P8, P11, P13, P15, P19).

\section{High-Level Codes}\label{app:dimensions}

Table~\ref{tab:dimensions} lists the 15 high-level codes under which the
merged low-level codes were organized, the number of low-level codes in each,
and the five most frequent low-level codes in each.

\begin{table}[t]
  \caption{High-level codes, with the number of low-level codes in each and
  the five most common.}
  \label{tab:dimensions}
  \footnotesize
  \begin{tabular}{@{}l r p{0.68\textwidth}@{}}
    \toprule
    High-level code & Low-level codes & Top five low-level codes \\
    \midrule
    Experience & 25 & open-source contributor or maintainer; industry or
    enterprise software; SDD used solo; works in teams; SDD used in a team with
    non-developer roles \\
    Adoption & 17 & motivation: agent context and memory limits; arrived at SDD
    organically; motivation: scaling beyond vibe coding; motivation: correctness
    and quality; introduced by an individual champion \\
    Spec content & 29 & goals, problem statement, and intent (what and why);
    design and architecture; constraints and guardrails; tasks and plans;
    requirements and user stories \\
    Spec structure & 27 & hierarchical decomposition (e.g., PRD or epic, spec,
    tasks); specs linked to each other and to code; specs scoped to individual
    changes; tool-defined template; specs not linked to each other or to code \\
    Spec workflow & 61 & spec-gated autonomous workflow; AI-authored specs; spec
    source: conversation or research with AI; task decomposition and
    orchestration; no SDD for small tasks, bugs, or one-offs \\
    Spec evaluation & 37 & tests verify code against spec; manual spec review;
    end-to-end or acceptance tests; multi-agent adversarial review;
    human-anchored verification (acceptance criteria or human-defined tests) \\
    Spec maintenance & 29 & specs committed with code; living spec kept current;
    specs as historical record and provenance; drift detection and checks; specs
    never deleted \\
    Tooling & 32 & built own SDD tool or framework; customization level: builds
    own tool or workflow; customizing SDD tools; compares or surveys SDD tools;
    uses OpenSpec \\
    Collaboration & 13 & collaboration as a goal rather than an observed outcome;
    team alignment on workflow; specs reviewed before code; modest team benefit,
    spec as shared review artifact; not applicable, SDD practiced solo \\
    Benefit & 26 & more reliable agent work; faster delivery and productivity;
    reduced tokens and rework; better context management; reduced review and
    cognitive effort \\
    Challenge & 64 & token and model costs; time overhead; spec-code drift; tool
    rigidity or heaviness; underspecification and unknown unknowns \\
    Learning curve & 13 & learned by doing and experimentation; advice for
    newcomers; practice evolved over time; steep learning curve; education and
    training needs \\
    Agentic SE & 30 & skills and plugins; context engineering and memory files;
    context window limits; agent failure modes (hallucination, shortcuts,
    overreach); model choice by task and cost \\
    SDD perception & 18 & relation to traditional RE and waterfall; planning is
    expensive, implementation cheap; SDD may be transitional as models improve;
    spec as memory and context for agents; intent must be clear before specifying \\
    Other & 6 & critique of software industry practices; project-specific
    technical detail; research interests beyond SDD; software and language design
    philosophy; personal context \\
    \midrule
    Total & 427 & \\
    \bottomrule
  \end{tabular}
\end{table}

\section{Codes and Participants}\label{app:codes}

\begin{longtable}{@{}r p{0.52\textwidth} p{0.36\textwidth}@{}}
  \caption{Participants assigned each code used in the findings (tool makers in bold). Codes are assigned independently, so a participant can carry a narrower code, such as a drift type, without the broader one. A participant absent from every code of a finding did not raise that topic in a way we coded.}\label{tab:codes-participants}\\
  \toprule
  Finding & Code & Participants \\
  \midrule
  \endfirsthead
  \toprule Finding & Code & Participants \\ \midrule \endhead
  \bottomrule \endfoot
  1 & spec-content: implementation detail included & P7, P12, \textbf{P13},
  P14, P15, P16, P18, P19, \textbf{P21} \\
  & spec-content: implementation detail included selectively & P4, P5, P8, P10 \\
  & spec-content: implementation detail kept out of spec & \textbf{P1},
  \textbf{P2}, P3, \textbf{P6}, \textbf{P9}, \textbf{P11}, P17, \textbf{P20} \\
  \midrule
  2 & spec-workflow: SDD use decided by task size or complexity & P4, P5, P10,
  \textbf{P11}, P12, \textbf{P13}, P14, P19 \\
  & spec-workflow: SDD use decided by risk or uncertainty & \textbf{P2}, P3,
  \textbf{P9}, P15, P17, \textbf{P20}, \textbf{P21} \\
  \midrule
  3 & spec-eval: full human spec review & \textbf{P2}, \textbf{P6},
  \textbf{P11}, P12, \textbf{P13}, P16, P19 \\
  & spec-eval: partial or cursory spec reading & P4, P5, P7, P8, P14, P15,
  \textbf{P20} \\
  & spec-eval: spec review primarily delegated to agents & P3, \textbf{P9},
  P10, P17, P18 \\
  & spec-eval: verification is hard or unsolved & \textbf{P1}, P3, P8, P10,
  P14, P16, P18, \textbf{P20} \\
  & spec-eval: human-anchored verification (acceptance criteria or
  human-defined tests) & \textbf{P1}, P4, P5, \textbf{P6}, \textbf{P9}, P12,
  \textbf{P13}, P16, P17, P19, \textbf{P21} \\
  \midrule
  4 & spec-maintenance: living spec kept current & \textbf{P1}, \textbf{P2},
  P3, \textbf{P6}, P8, \textbf{P9}, P16, P18, P19, \textbf{P20} \\
  & spec-maintenance: specs kept as accumulated record & P4, P5, P7, P10,
  \textbf{P11}, \textbf{P13} \\
  & spec-maintenance: specs temporary, discarded after implementation & P12,
  P14, \textbf{P21} \\
  \midrule
  5 & tooling: customization level: uses defaults & P5, P15 \\
  & tooling: customization level: extends an existing tool & P4, P7, P8, P10,
  P12, P16, P17, P19 \\
  & tooling: customization level: builds own tool or workflow & \textbf{P1},
  \textbf{P2}, P3, \textbf{P6}, \textbf{P9}, \textbf{P11}, \textbf{P13}, P14,
  P18, \textbf{P20}, \textbf{P21} \\
  & tooling: customizes for organizational fit & P4, \textbf{P6}, P10, P12, P19 \\
  & tooling: switched to a lighter or more customizable tool & P7, P8,
  \textbf{P11}, \textbf{P13}, P15, P19 \\
  & spec-structure: specs scoped to system capabilities or subsystems &
  \textbf{P1}, \textbf{P2}, P3, \textbf{P6}, P7, P14, \textbf{P20} \\
  & spec-structure: specs scoped to individual changes & P4, P5, P8, P10,
  \textbf{P11}, P12, \textbf{P13}, P15, P16, P17, P18, P19, \textbf{P21} \\
  & spec-structure: specs linked to each other and to code & \textbf{P2}, P4,
  P5, \textbf{P6}, P8, \textbf{P9}, \textbf{P11}, \textbf{P13}, P14, P15, P16,
  P18, \textbf{P20} \\
  & spec-structure: specs not linked to each other or to code & \textbf{P1},
  P3, P7, P10, P12, P17, P19, \textbf{P21} \\
  \midrule
  6 & collaboration: shared understanding across roles & \textbf{P6}, P10, P16 \\
  & collaboration: modest team benefit, spec as shared review artifact & P4,
  P5, \textbf{P13}, P19 \\
  & collaboration: little or no change in human collaboration & P12, P14, P17 \\
  & collaboration: collaboration as a goal rather than an observed outcome &
  \textbf{P1}, \textbf{P2}, P3, \textbf{P9}, \textbf{P11}, \textbf{P20}, \textbf{P21} \\
  \midrule
  7 & benefit: more reliable agent work & \textbf{P1}, \textbf{P2}, P4, P5,
  \textbf{P6}, P7, P8, \textbf{P9}, P10, \textbf{P11}, \textbf{P13}, P14, P15,
  P16, P17, P19, \textbf{P20}, \textbf{P21} \\
  \midrule
  8 & benefit: clarity of thinking and intent & \textbf{P1}, \textbf{P9}, P18, P19 \\
  & benefit: agent-written spec exposes gaps in own intent & P7, \textbf{P11},
  P14, P17, \textbf{P20} \\
  & sdd-perception: intent must be clear before specifying & \textbf{P1},
  \textbf{P2}, \textbf{P9}, P17, \textbf{P21} \\
  \midrule
  9 & challenge: token and model costs & \textbf{P1}, P3, P5, \textbf{P6}, P7,
  P8, P10, \textbf{P13}, P14, P15, P16, P17, P18, P19 \\
  & challenge: planning consumes most tokens & P3, P10, P18, P19 \\
  & benefit: reduced tokens and rework & P4, \textbf{P9}, \textbf{P11},
  \textbf{P13}, P14, P16, P17 \\
  & challenge: time overhead & \textbf{P1}, P3, P4, \textbf{P6}, P10,
  \textbf{P13}, P16, P17, P18, \textbf{P21} \\
  & benefit: faster delivery and productivity & \textbf{P6}, P8, P10,
  \textbf{P11}, P14, P16, P19 \\
  & benefit: reduced review and cognitive effort & \textbf{P2}, P10,
  \textbf{P11}, \textbf{P13}, \textbf{P20} \\
  \midrule
  10 & challenge: spec-code drift & \textbf{P2}, P5, \textbf{P6}, P7, P8, P10,
  \textbf{P11}, P12, \textbf{P13}, P14, P16, P18, P19, \textbf{P21} \\
  & challenge: drift type: agent deviates during implementation & P5,
  \textbf{P6}, P7, P8, P16, P17, P18, P19 \\
  & challenge: drift type: spec goes stale as code evolves & \textbf{P2}, P8,
  P10, \textbf{P11}, P12, P14, \textbf{P21} \\
  & spec-maintenance: drift detection and checks & \textbf{P2}, \textbf{P6},
  P8, P10, \textbf{P11}, P12, P14 \\
  \midrule
  11 & challenge: spec correctness hard to evaluate & P8, P10 \\
  & challenge: increased review load and large PRs & P4, P8, P10, \textbf{P11},
  P12, P14 \\
  & challenge: spec corpus grows like a codebase & P4, P10, \textbf{P11}, P17, P18 \\
  & challenge: cross-spec dependencies and complex features & \textbf{P2},
  P4, P10, P18 \\
  & challenge: difficulties as project size grows & P3, P4, P14, P15, P18 \\
  \midrule
  12 & challenge: organizational adoption inertia & P5, \textbf{P6},
  \textbf{P9}, P19, \textbf{P21} \\
  & adoption: introduced by an individual champion & P3, P4, P5, \textbf{P6},
  \textbf{P9}, \textbf{P13}, \textbf{P21} \\
  & adoption: adopted as an organizational process & P10, P12, P16, P19 \\
  & adoption: fits an existing formal process & P10, P12 \\
\end{longtable}

\end{document}